\documentclass[11pt]{article}

\usepackage[final]{acl}
\usepackage{times}
\usepackage{latexsym}
\usepackage[T1]{fontenc}
\usepackage[utf8]{inputenc}
\usepackage{microtype}
\usepackage{inconsolata}
\usepackage{graphicx}
\usepackage{booktabs}
\usepackage{amsmath}
\usepackage{amssymb}
\usepackage{pifont}
\usepackage{multirow}

\title{ProTTS: End-to-End Prominence-aware Natural and Controllable Text-to-Speech via Automatic Word Stress Discovery}

\author{Sai Harshitha Aluru \\
    Speech Processing Lab, LTRC \\
    IIIT Hyderabad \\
    Hyderabad, India \\
  \texttt{saiharshitha.aluru@research.iiit.ac.in} \\\And
  Chiranjeevi Yarra \\
    Speech Processing Lab, LTRC \\
    IIIT Hyderabad \\
    Hyderabad, India \\
  \texttt{chiranjeevi.yarra@iiit.ac.in} \\}

\begin{document}

\maketitle

\begin{abstract}
Natural speech conveys meaning not only through words, but also through the prominence on a few words. Yet most text-to-speech (TTS) systems remain prominence-unaware, existing prominence-aware TTS require external stress annotations, while unsupervised stress detectors operate in isolation, disconnected from synthesis. This creates a fundamental gap no existing system can automatically discover where stress should be placed, realize it as natural-sounding prosody, and offer user control over emphasis placement, all within a single model. To circumvent this, we propose ProTTS, a unified framework where stress detection emerges from the TTS model itself and, in turn, guides it to synthesize more natural and controllable speech. We incorporate a clustering head into the variance adaptor of FastSpeech\,2 that discovers word stress directly from aggregated prosodic features. The discovered stress then drives emphasis variance predictors that modify pitch, energy, and duration for stressed words. The framework is trained end-to-end: TTS reconstruction and clustering losses jointly shape a stress-discriminative feature space without requiring stress labels at any stage. At inference, the same architecture also supports user-controllable emphasis by accepting a custom stress mask. We evaluate ProTTS on ISLE (non-native), LibriTTS, VCTK (native training), and three evaluation benchmarks (Tatoeba, TinyStress-15k, EmphAssess). Listeners prefer ProTTS over the baseline with a MOS of 3.96 vs.\ 3.12 and 83.5\% preference rate. For stress detection, ProTTS achieves 0.839/0.861 F1 on ISLE (German/Italian L2), outperforming unsupervised ($\sim$0.70) and supervised ($\sim$0.82) baselines, and 0.933 F1 in zero-shot transfer to Tatoeba.
\end{abstract}

\section{Introduction}
\label{sec:intro}

In spoken language, certain words are produced with greater intensity, longer duration, and pitch variation (known as stress) to convey communicative intent. Listeners rely on this word-level stress to interpret meaning, resolve ambiguity, and perceive naturalness in speech \citep{terken2000perception, cutler1997prosody, vanheuven2018acoustic, cole2019sound}. Accurate stress placement is therefore especially important in expressive contexts, where the same sentence can convey correction, contrast, or surprise depending on which word carries emphasis. For second-language (L2) learners, producing and perceiving correct stress patterns remains a persistent challenge \citep{cutler1997prosody}. Stress-aware speech synthesis therefore has broad applications from producing more natural voice assistants, to enabling controllable emphasis in audiobook narration, to providing corrective feedback in Computer-Assisted Language Learning (CALL) systems \citep{gasteiger2024scoping}.

Despite its importance, word-level stress remains largely absent from modern text-to-speech systems. Even state-of-the-art models fail to reliably produce contextually appropriate stress \citep{cast2026}. Research on stress detection and emphasis-controllable synthesis has progressed along independent tracks. Stress detection methods, whether supervised or unsupervised, are trained and evaluated in isolation without considering how detected stress might be realized in speech. Emphasis-controllable TTS models can modulate prosody for designated words, but rely on externally provided stress labels, assuming the problem of \emph{where} to stress has already been solved. This leaves a disconnect between knowing \textit{where to stress} and \textit{how to produce it}.

A straightforward solution would be to cascade a stress detector with an emphasis-controllable synthesizer, as in \citet{li2025emetss}. However, an external detector optimized on its own objective produces stress labels that may not reflect the prosodic distinctions that are useful for synthesis. At the same time, the TTS model's own predictions (pitch, energy, and duration) already encode acoustic contrasts between stressed and unstressed words, yet this signal is never exploited for stress discovery in a cascaded setup. This raises a key question: \emph{\textit{can we design a system where stress detection and stress-aware synthesis are mutually beneficial, one that automatically discovers emphasis, produces natural speech, and supports controllable stress placement?}}

To address this, we propose ProTTS, a unified framework where stress detection and stress-aware synthesis emerge jointly from a single model. Our key insight is that the prosodic features predicted inside a TTS variance adaptor (pitch, energy, and duration) are inherently stress-discriminative. Stressed words naturally exhibit different prosodic profiles than unstressed ones, and this distinction can be discovered without supervision. Motivated by this observation, we integrate a clustering head within the variance adaptor of FastSpeech\,2 that identifies word stress directly from aggregated prosodic features, requiring no external labels. The discovered stress then drives emphasis variance predictors that learn how pitch, energy, and duration should change for stressed words; unstressed words retain their base prosody unchanged. The framework is trained end-to-end with a joint objective combining TTS reconstruction and stress-clustering losses. At inference, the same architecture also supports user-controllable emphasis by accepting a custom stress mask in place of the clustering head's predictions. We evaluate across both native and non-native speech settings.

Our contributions are as follows: \\
(1) We propose the first framework that unifies unsupervised word stress detection and stress-aware synthesis within a single TTS model, trained end-to-end without any stress annotations. The same architecture also supports user-controllable emphasis at inference.\\
(2) We demonstrate strong stress detection performance across both native and non-native speech, outperforming supervised baselines despite using no labels, with zero-shot transfer to unseen evaluation sets.\\
(3) Through subjective evaluations, we show that listeners reliably perceive the intended stress and consistently prefer ProTTS's synthesis over the baseline.

Additionally, we present ablation studies on feature composition, joint versus cascaded training, cross-domain transfer from native to non-native speech, and an analysis of the learned prosodic representations.
\section{Related Work}
\label{sec:related}

\subsection{Word Stress Detection}

Early approaches to word stress detection relied on handcrafted prosodic features (fundamental frequency, energy, duration, and spectral tilt) combined with conventional classifiers \citep{kakouros2015automatic, kakouros2017spectral, sabu2018detection}, HMM-based frameworks \citep{li2007hmm}, or sequential models such as bidirectional RNNs \citep{rosenberg2015modeling}. More recently, self-supervised speech models have reduced the dependence on manual feature engineering. Representations from wav2vec\,2.0 \citep{baevski2020wav2vec}, HuBERT \citep{hsu2021hubert}, and Whisper \citep{radford2023whisper} have been applied to stress. \citet{yosha2025whistress} predicts word-level stress from Whisper representations, while \citet{mallela2025supradoral} models suprasegmental dependencies for detecting the same. These methods demonstrate that pretrained representations contain rich prosodic information, but they are predominantly supervised, requiring word-level stress annotations that are expensive to obtain, particularly for non-native speech where stress patterns vary across L1 backgrounds. \\
Unsupervised stress detection remains comparatively underexplored. \citet{mondal2024preliminary} conduct a comparative analysis of FastSpeech\,2 prosody embeddings for prominence detection, showing that K-Means clustering on duration, energy, and pitch embeddings can detect word stress without supervision, though with substantially lower F1 than supervised methods.

\subsection{Emphasis-Controllable Speech Synthesis}

Emphasis control in non-autoregressive TTS typically involves modifying the variance adaptor's pitch, energy, and duration predictions for designated words \citep{ren2021fastspeech2}. \citet{li2025emetss} trains emphasis control alongside emotional expressivity using weakly supervised pseudo-labels. \citet{mondal2025exagtts} incorporates word stress into GlowTTS for L2 learners using a separately trained classifier. \citet{chen2025stresstransfer} preserves emphasis across speech-to-speech translation via LLM-based cross-lingual stress mapping. \citet{joly2023controllable} show that simply scaling phoneme duration produces effective emphasis, though limited to duration manipulation. \\
Across all these methods, stress labels must be provided externally. None integrates stress discovery into the synthesis model itself, meaning the two tasks cannot benefit from shared representations or joint optimization.

\subsection{Unsupervised Prosody Modeling in TTS}

Several lines of work learn prosodic structure within TTS models without explicit labels. Global Style Tokens \citep{wang2018style} learn soft style clusters for utterance-level speaking style control. This has been extended to finer granularities: \citet{chien2021hierarchical} propose hierarchical prosody modeling conditioned on word-level features, and \citet{guo2022unsupervised} introduce unsupervised word-level prosody tagging for controllable synthesis. However, these approaches model general prosodic variation (speaking rate, intonation contour, expressiveness) rather than the specific linguistic phenomenon of word stress. \\
Word stress is a discrete, binary property determined by both acoustic realization and communicative intent \citep{cole2019sound, kakouros2016analyzing}. To the best of our knowledge, no prior work has attempted to discover word stress unsupervised from within a TTS model and simultaneously use it to improve synthesis.

\section{Method}
\label{sec:method}

\begin{figure*}[t]
    \centering
    \includegraphics[width=\textwidth]{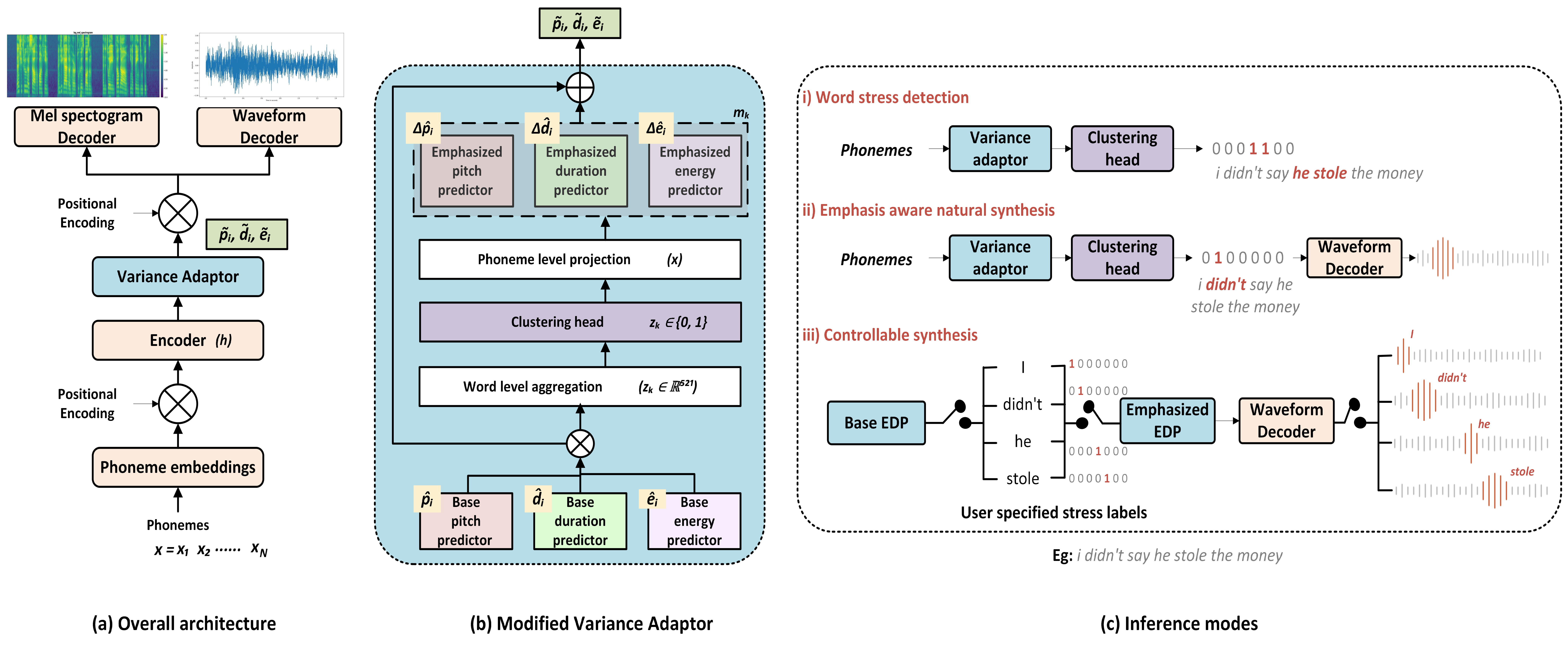}
    \caption{Architecture of ProTTS. (a) Overall FastSpeech\,2 model with our modified variance adaptor. (b) Inside the modified variance adaptor: base prosody predictions are aggregated into 521-dim word features, a clustering head discovers stressed words, and emphasis variance predictors apply learned deltas only to stressed words, gated by $m_k$. (c) Three inference modes: (i) word stress detection outputs binary labels without audio generation, (ii) emphasis-aware synthesis automatically discovers and realizes stress, and (iii) controllable synthesis allows users to specify stress placement, producing different prosodic realizations and meanings from the same sentence.}
    \label{fig:architecture}
\end{figure*}

ProTTS integrates unsupervised word stress detection directly into the variance adaptor of FastSpeech\,2, enabling joint stress discovery and stress-aware synthesis within a single model. Figure~\ref{fig:architecture}(a) shows the overall architecture, where the standard variance adaptor is replaced with our modified version containing a clustering head and emphasis variance predictors. Figure~\ref{fig:architecture}(b) details the modified variance adaptor's internal structure, and Figure~\ref{fig:architecture}(c) illustrates the three inference modes supported by the framework. We describe each component below.

\subsection{Background: FastSpeech\,2}

FastSpeech\,2 \citep{ren2021fastspeech2} is a non-autoregressive TTS model comprising a phoneme encoder, a variance adaptor, and a mel-spectrogram decoder. Given a phoneme sequence $\mathbf{x} = (x_1, \ldots, x_N)$, the encoder produces hidden representations $\mathbf{h} = (h_1, \ldots, h_N)$. The variance adaptor predicts three prosodic attributes at the phoneme level: pitch $\hat{p}_i$, energy $\hat{e}_i$, and duration $\hat{d}_i$. Pitch and energy values are passed through learned embedding layers to produce $\mathbf{p}_i^{\text{emb}} \in \mathbb{R}^{256}$ and $\mathbf{e}_i^{\text{emb}} \in \mathbb{R}^{256}$, which are added to the hidden sequence. A length regulator expands the sequence according to predicted durations, and the decoder generates the mel-spectrogram.

In the standard model, the variance adaptor treats all phonemes uniformly, it has no notion of which words should receive prosodic prominence. ProTTS modifies it to (1) aggregate phoneme-level predictions into word-level representations, (2) discover stressed words via a clustering head, and (3) apply learned prosodic shifts through emphasis variance predictors.

\subsection{Word-Level Prosodic Feature Aggregation}
\label{sec:aggregation}

Since stress is a word-level phenomenon but the variance adaptor operates at the phoneme level, we bridge this gap by aggregating the variance adaptor's own prosodic predictions into word-level representations (Figure~\ref{fig:architecture}(b)). This ensures the clustering head operates in a space that is both prosodically informative and end-to-end differentiable with respect to the TTS reconstruction loss.

Let word $w_k$ span phonemes $\{x_{s_k}, \ldots, x_{e_k}\}$. We construct a word-level feature vector $\mathbf{z}_k \in \mathbb{R}^{521}$ by mean-pooling the phoneme-level pitch embeddings $\mathbf{p}_i^{\text{emb}} \in \mathbb{R}^{256}$ and energy embeddings $\mathbf{e}_i^{\text{emb}} \in \mathbb{R}^{256}$ within the word. Since duration predictions in FastSpeech\,2 are scalar values rather than learned embeddings, we aggregate them into a fixed-length word-level representation using summary statistics (mean, std, max, min, total, range, coefficient of variation), along with normalized word position ($\text{pos}_k$) and phoneme count ($n_k^{\text{ph}}$):
\begin{equation}
    \mathbf{z}_k = [\mathbf{z}_k^{(p)} \;\Vert\; \mathbf{z}_k^{(e)} \;\Vert\; \mathbf{z}_k^{(d)}] \in \mathbb{R}^{521}
    \label{eq:train_feats}
\end{equation}
where $\mathbf{z}_k^{(p)} \in \mathbb{R}^{256}$, $\mathbf{z}_k^{(e)} \in \mathbb{R}^{256}$, and $\mathbf{z}_k^{(d)} \in \mathbb{R}^{9}$. Detailed equations are in Appendix~\ref{app:features}.

\subsection{Clustering Head for Stress Discovery}
\label{sec:clustering}

Rather than training a separate classifier, which would require stress labels, we discover the stressed/unstressed distinction directly from the prosodic feature space using a clustering approach inspired by deep embedded clustering \citep{guo2017idec}. The clustering head partitions word-level representations into two categories through soft assignment.

Given $K=2$ learnable centroids $\{\boldsymbol{\mu}_0, \boldsymbol{\mu}_1\} \in \mathbb{R}^{521}$, we compute a soft assignment $q_{kj}$ for each word $\mathbf{z}_k$ using a Student's $t$-distribution kernel:
\begin{equation}
    q_{kj} = \frac{(1 + \lVert\mathbf{z}_k - \boldsymbol{\mu}_j\rVert^2)^{-1}}{\sum_{j'=0}^{1} (1 + \lVert\mathbf{z}_k - \boldsymbol{\mu}_{j'}\rVert^2)^{-1}}
    \label{eq:soft_assign}
\end{equation}

To encourage sharper assignments, we define a target distribution $P$ by squaring \& normalizing $Q$:
\begin{equation}
    p_{kj} = \frac{q_{kj}^2 / f_j}{\sum_{j'} q_{kj'}^2 / f_{j'}}
    \label{eq:target}
\end{equation}
where $f_j = \sum_k q_{kj}$ is the soft cluster frequency, preventing large clusters from dominating. The clustering loss minimizes KL divergence between $P$ and $Q$:
\begin{equation}
    \mathcal{L}_{\text{kl}} = \text{KL}(P \| Q) = \sum_k \sum_j p_{kj} \log \frac{p_{kj}}{q_{kj}}
    \label{eq:kl_loss}
\end{equation}

Since $P$ is derived from $Q$ itself, this creates a self-reinforcing loop that progressively sharpens the cluster boundaries, discovering stressed words without supervision.

\textbf{Centroid initialization.} $\boldsymbol{\mu}_0$ and $\boldsymbol{\mu}_1$ are initialized via K-Means on the pretrained FastSpeech\,2 prosodic features, following \citep{guo2017idec}. We use only duration and pitch features for initialization, as they yield better separation than the full EDP combination (Section~\ref{sec:ablations}).

\textbf{Hard assignment.} At each epoch boundary, hard pseudo-labels are derived from the soft assignments:
\begin{equation}
    y_k = \underset{j}{\operatorname{argmax}} \; q_{kj}
    \label{eq:hard_assign}
\end{equation}
These serve as the emphasis mask $m_k$ for the variance predictors (Section~\ref{sec:emphasis}) and determine which words receive variance supervision at each epoch.

\textbf{Cluster identity assignment.} Since the two clusters have no inherent label, we assign identity using a prosodic prior: the centroid with the larger L2 norm is designated as the stressed cluster. Stressed words exhibit higher pitch, greater energy, and longer duration \citep{vanheuven2018acoustic}, producing embeddings with larger magnitudes and a centroid further from the origin.

\subsection{Emphasis-Gated Variance Adaptation}
\label{sec:emphasis}

Following \citet{li2025emetss}, we augment the variance adaptor with three \emph{emphasis variance predictors} (Figure~\ref{fig:architecture}(b)), one each for pitch, energy, and duration. Given the emphasis mask $m_k \in \{0, 1\}$ from the clustering head, the emphasis predictors produce phoneme-level deltas that are added to the \emph{base prosody} ($\hat{p}_i$, $\hat{e}_i$, $\hat{d}_i$) to yield the \emph{stress-adapted prosody}:
\begin{align}
    \tilde{p}_i &= \hat{p}_i + \Delta \hat{p}_i \cdot m_k \label{eq:pitch_final} \\
    \tilde{e}_i &= \hat{e}_i + \Delta \hat{e}_i \cdot m_k \label{eq:energy_final} \\
    \tilde{d}_i &= \hat{d}_i + \Delta \hat{d}_i \cdot m_k \label{eq:dur_final}
\end{align}

When $m_k = 0$ (unstressed), the deltas vanish and the phoneme retains its base prosody. When $m_k = 1$ (stressed), the learned deltas produce the elevated pitch, increased energy, and longer duration characteristic of stressed words \citep{vanheuven2018acoustic}. The stress-adapted prosody is passed to the length regulator and decoder for mel-spectrogram generation.

\subsection{Joint Training Objective}
\label{sec:training}

All components are differentiable and trained jointly. The full objective is:
\begin{equation}
    \mathcal{L}_{\text{total}} = \mathcal{L}_{\text{recon}} + \alpha \cdot (\mathcal{L}_{\text{kl}} + \mathcal{L}_{\text{bal}})
    \label{eq:total_loss}
\end{equation}

\textbf{Reconstruction loss.} Following \citet{li2025emetss}, $\mathcal{L}_{\text{recon}}$ comprises mel-spectrogram losses and prosody prediction losses:
\begin{equation}
\begin{split}
    \mathcal{L}_{\text{recon}} = & \; \mathcal{L}_{\text{mel}} + \mathcal{L}_{\text{postnet}} + \mathcal{L}_{\text{pitch}} + \mathcal{L}_{\text{energy}} \\
    & + \mathcal{L}_{\text{dur}} + \mathcal{L}_{\text{pv}} + \mathcal{L}_{\text{dv}} + \mathcal{L}_{\text{ev}}
\end{split}
    \label{eq:recon_loss}
\end{equation}
The \emph{combined prosody losses} supervise the stress-adapted prosody against ground-truth targets, e.g.\ $\mathcal{L}_{\text{pitch}} = \text{MSE}(\hat{p}_i + \Delta \hat{p}_i,\; p_i^{\text{gt}})$, with energy and duration defined analogously. The \emph{variance losses} separately supervise the emphasis deltas against precomputed variance targets, e.g.\ $\mathcal{L}_{\text{pv}} = \text{MSE}(\Delta \hat{p}_i,\; \Delta p^{\text{gt}})$. Following a similar strategy to \citet{li2025emetss}, the variance targets $\Delta p^{\text{gt}}$, $\Delta d^{\text{gt}}$, $\Delta e^{\text{gt}}$ are computed as the difference between the average prosody of emphasized phonemes and the sentence-level average, and are regenerated each epoch as pseudo-labels evolve. Full loss definitions and variance target formulations are in Appendix~\ref{app:losses}.

\textbf{Clustering loss.} $\mathcal{L}_{\text{kl}}$ is the KL divergence from Eq.~\ref{eq:kl_loss}. $\mathcal{L}_{\text{bal}}$ penalizes degenerate solutions where all words collapse into one cluster:
\begin{equation}
    \mathcal{L}_{\text{bal}} = \left(\frac{f_0}{f_0 + f_1} - 0.5\right)^2
    \label{eq:balance_loss}
\end{equation}
$\alpha$ balances synthesis quality against stress discriminability.

\textbf{Training procedure.} The model is initialized from a pretrained FastSpeech\,2 checkpoint. After each epoch, pseudo-labels and variance targets are regenerated from the updated soft assignments, creating a feedback loop that progressively refines both stress detection and synthesis. Full details are in Appendix~\ref{app:training}.

\subsection{Inference}
\label{sec:inference}

The trained model supports three inference modes (Figure~\ref{fig:architecture}(c)). In \textbf{word stress detection} mode (Figure~\ref{fig:architecture}(c-i)), the clustering head outputs binary stress labels from the aggregated prosodic features without generating audio, enabling the framework to function as a standalone stress detector. In \textbf{emphasis-aware synthesis} mode (Figure~\ref{fig:architecture}(c-ii)), the discovered stress mask gates the emphasis variance predictors to produce stress-adapted prosody, and the decoder generates the mel-spectrogram. A HiFi-GAN vocoder converts the output to a waveform. In \textbf{controllable synthesis} mode (Figure~\ref{fig:architecture}(c-iii)), the emphasis mask $m_k$ is replaced with user-specified stress labels, bypassing the clustering head and allowing explicit control over which words receive prosodic prominence without retraining. As illustrated, placing stress on different words in the same sentence (e.g., \emph{``\textbf{I} didn't say he stole the money''} vs.\ \emph{``I \textbf{didn't} say he stole the money''}) produces distinct prosodic realizations that convey different meanings. Additional controllability examples are provided in Appendix~\ref{app:synthesis_breakdown}.

\section{Experiments and Results}
\label{sec:experiments}

\subsection{Experimental Setup}
\label{sec:setup}

We design experiments to answer four questions: (1) Can ProTTS discover word stress without labels, and how close does it come to supervised methods? (2) Does stress learned from one domain transfer to another? (3) Does the discovered stress translate into perceptibly better synthesis? (4) How do key design choices affect performance?

\textbf{Datasets.} We conduct experiments along two tracks. \emph{Track 1 (Non-native):} We train and evaluate on the ISLE corpus \citep{menzel2000isle}, containing L2 English speech from German (GER) and Italian (ITA) speakers. Word-level stress annotations were obtained manually (Appendix~\ref{app:annotation}) and used only for evaluation; the model never sees them during training. \emph{Track 2 (Native):} We train separately on LibriTTS \citep{zen2019libritts} and VCTK \citep{veaux2017vctk}, both native English corpora without stress annotations. Stress detection is evaluated on three test sets: Tatoeba \citep{tatoeba} (manually annotated, Appendix~\ref{app:annotation}), TinyStress-15k \citep{yosha2025whistress}, and EmphAssess \citep{deseyssel2024emphassess}. For synthesis evaluation, we use the LibriTTS and VCTK test sets. Dataset statistics are in Appendix~\ref{app:datasets}.

\textbf{Baselines.} For stress detection: unsupervised K-Means clustering on FastSpeech\,2 prosody embeddings \citep{mondal2024preliminary}, and supervised methods including a DNN on the same embeddings \citep{mondal2024preliminary}, SupraDoRAL \citep{mallela2025supradoral}, and WhiStress \citep{yosha2025whistress}. For synthesis: base FastSpeech\,2 without any emphasis mechanism.

\textbf{Evaluation protocol.} \emph{Stress detection:} Word-level F1 against ground-truth annotations, reported per L1 background (GER, ITA) on ISLE and per native test set.
\emph{Synthesis quality:} Three subjective evaluations with 35 listeners of diverse backgrounds, balanced across gender, native language, and familiarity with speech research (Appendix~\ref{app:listening_test}): (a) \emph{MOS + Preference}: listeners rate two audio samples on a 1--5 naturalness scale and select the more natural one; (b) \emph{Stress preference}: listeners see target stressed words highlighted and select which audio better realizes the stress (with ``both same'' and ``neither good'' options); (c) \emph{ABX similarity}: listeners hear an original reference and two generated versions, and select which is closer to the original.

\textbf{Training.} We initialize from a pretrained FastSpeech\,2 checkpoint and train for 100k steps with $\alpha = 1.0$. Full hyperparameters and convergence analysis are in Appendices~\ref{app:hyperparams} and~\ref{app:convergence}.

\textbf{Ablation configurations.} We evaluate three design choices:
(1) \emph{Joint vs.\ cascaded.} We compare our end-to-end framework against a cascaded baseline where FS2 and a DNN stress classifier are trained in alternation but never jointly. FS2 is first fine-tuned without stress, then prosodic embeddings are clustered via K-Means, a DNN is trained on these labels, and DNN predictions (confidence $> 0.8$, else K-Means fallback) serve as stress labels for the next FS2 iteration. Details in Appendix~\ref{app:cascaded}.
(2) \emph{Feature composition.} Clustering on energy+duration+pitch (EDP) vs.\ duration+pitch only (DP) at centroid initialization.
(3) \emph{Cross-domain transfer.} Training on native speech (LibriTTS/VCTK) and evaluating on non-native speakers (ISLE).

\subsection{Results and Discussion}
\label{sec:results}

\subsubsection{Word Stress Detection}

\begin{table}[t]
\centering
\resizebox{\columnwidth}{!}{
\begin{tabular}{@{}llcc@{}}
\toprule
& \textbf{Method} & \textbf{GER} & \textbf{ITA} \\
\midrule
\multirow{4}{*}{Sup.$^\dagger$} & DNN \citep{mondal2024preliminary} & 0.83 & 0.74 \\
& SupraDoRAL \citep{mallela2025supradoral} & 0.78 & 0.87 \\
& WhiStress \citep{yosha2025whistress} & 0.80 & 0.87 \\
& \textbf{ProTTS (sup.)} & \textbf{0.901} & \textbf{0.932} \\
\midrule
\multirow{2}{*}{Unsup.} & K-Means \citep{mondal2024preliminary} & 0.76 & 0.71 \\
& \textbf{ProTTS (unsup.)} & \textbf{0.839} & \textbf{0.861} \\
\bottomrule
\end{tabular}
}
\vspace{-0.2cm}
\caption{Word stress detection F1 scores on ISLE (non-native). $\dagger$: supervised (requires stress labels)}
\label{tab:stress_nonnative}
\vspace{-0.5cm}
\end{table}

\begin{table}[t]
\centering
\resizebox{\columnwidth}{!}{
\begin{tabular}{@{}llccc@{}}
\toprule
& \textbf{Method} & \textbf{Tat.} & \textbf{Tiny.} & \textbf{Emph.} \\
\midrule
\multirow{2}{*}{Sup.$^\dagger$} & DNN \citep{mondal2024preliminary} & 0.891 & --- & --- \\
& WhiStress \citep{yosha2025whistress} & --- & 0.901 & 0.93 \\
\midrule
Unsup. & K-Means \citep{mondal2024preliminary} & 0.71 & --- & --- \\
\midrule
\multirow{2}{*}{\textbf{ProTTS}} & \textbf{unsup.}$^a$ & \textbf{0.933} & \textbf{0.911} & 0.923 \\
& unsup.$^b$ & 0.921 & 0.90 & 0.915 \\
\bottomrule
\end{tabular}
}
\vspace{-0.2cm}
\caption{Word stress detection F1 scores on native datasets. $\dagger$: supervised. Tat.=Tatoeba, Tiny.=TinyStress-15k, Emph.=EmphAssess. $^a$Trained on LibriTTS. $^b$Trained on VCTK.}
\label{tab:stress_native}
\vspace{-0.5cm}
\end{table}

Tables~\ref{tab:stress_nonnative} and~\ref{tab:stress_native} present word stress detection results on native and non-native datasets.

\textbf{Native (Table~\ref{tab:stress_native}).} On Tatoeba, training on LibriTTS yields 0.933 F1, surpassing both the K-Means baseline (0.71) and the supervised DNN (0.891). VCTK achieves 0.921. \textit{This addresses our second question: stress patterns learned from one corpus generalize to unseen data, and the clustering head is not memorizing corpus-specific artifacts but capturing language-level regularities in how stress is realized prosodically.} On TinyStress-15k and EmphAssess, our LibriTTS-trained model achieves 0.911 and 0.923 respectively, competitive with or exceeding the supervised WhiStress baseline (0.901 and 0.93).

\textbf{Non-native (Table~\ref{tab:stress_nonnative}).} ProTTS achieves 0.839 and 0.861 F1 for German and Italian L2 speakers, surpassing not only the unsupervised K-Means baseline (0.76/0.71) but, remarkably, all supervised methods as well: DNN (0.83/0.74), SupraDoRAL (0.78/0.87), and WhiStress (0.80/0.87). \textit{This directly addresses our first question: ProTTS discovers word stress without any labels, and does so better than methods that have access to them.} The higher F1 on Italian speakers aligns with the observation that Italian L2 English tends to produce more exaggerated prosodic contrasts on stressed words \citep{menzel2000isle}, while these patterns differ from native English stress realization, the exaggeration makes the stressed/unstressed boundary more acoustically salient and easier for the clustering head to discover in the in-domain setting. To establish a supervised upper bound, we train a DNN stress classifier on ground-truth labels using the same cascaded setup as our ablation (Appendix~\ref{app:cascaded}), where the DNN's predictions serve as the emphasis mask. At inference, the DNN achieves 0.901/0.932 F1. The remaining gap of 0.06--0.07 from our unsupervised results likely reflects cases where stress depends on discourse context or speaker intent rather than acoustic cues alone. Native speech yields higher F1 than non-native overall, which is expected, native speakers tend to produce more consistent stress patterns, giving the clustering head a cleaner signal to work with.

\begin{figure}[t]
    \centering
    \includegraphics[width=\columnwidth]{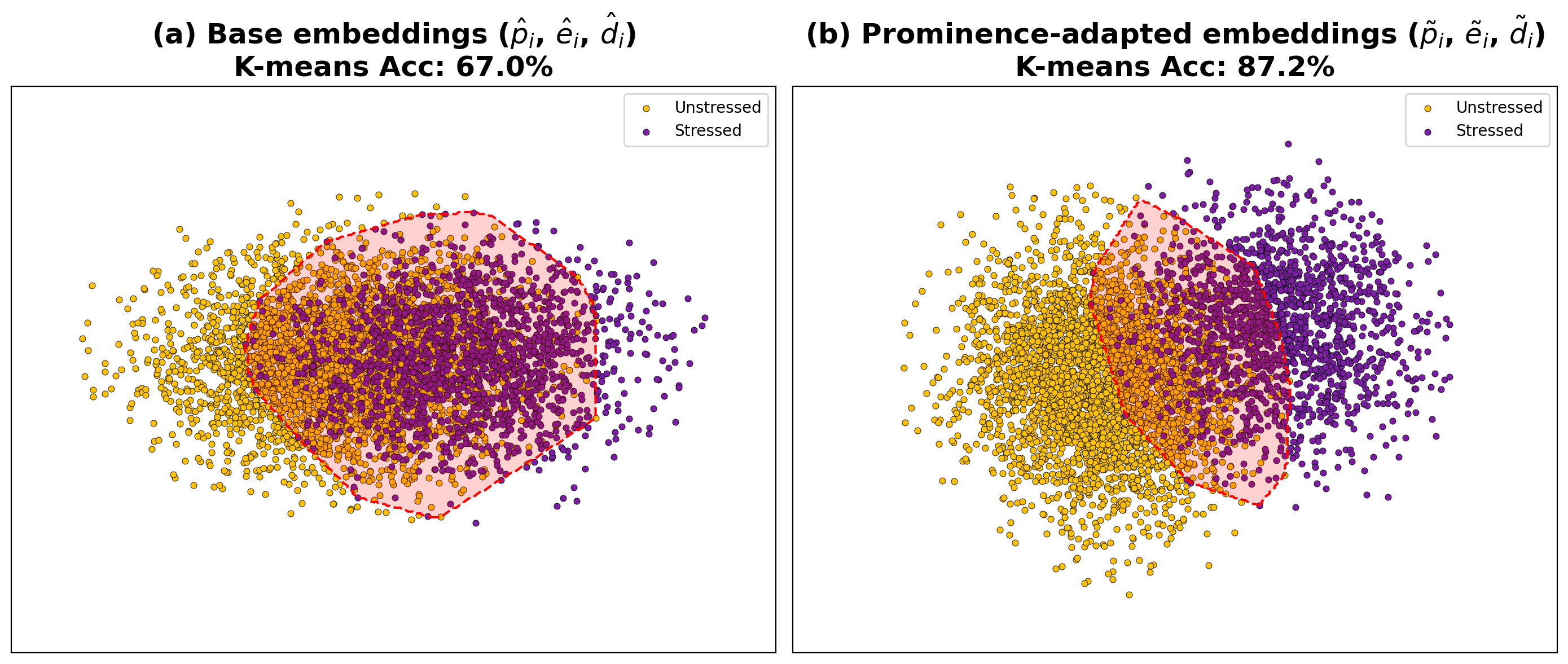}
    \vspace{-0.5cm}
    \caption{PCA of word-level EDP embeddings on ISLE, colored by ground-truth stress. Left: base prosody embeddings ($\hat{p}_i$, $\hat{e}_i$, $\hat{d}_i$). Right: stress-adapted prosody embeddings ($\tilde{p}_i$, $\tilde{e}_i$, $\tilde{d}_i$).}
    \label{fig:pca}
    \vspace{-0.2cm}
\end{figure}

\textbf{Learned prosodic representations.} We attribute ProTTS's strong performance to the emphasis-gated design. Because the variance predictors act exclusively on words identified as stressed, the model is incentivized to produce prosodic features that cleanly separate stressed from unstressed words. Over training, this creates a virtuous cycle: better cluster separation leads to more accurate emphasis masks, which produce more targeted prosodic shifts, which in turn further sharpens the feature space. Figure~\ref{fig:pca} provides direct evidence: K-Means on base prosodic embeddings before training achieves only 67.0\% clustering accuracy, while K-Means on stress-adapted embeddings after joint training achieves 87.2\%, confirming that the model learns stress-discriminative representations without ever being told which words are stressed.

\subsubsection{Synthesis Quality}

\begin{figure}[t]
    \centering
    \includegraphics[width=\columnwidth]{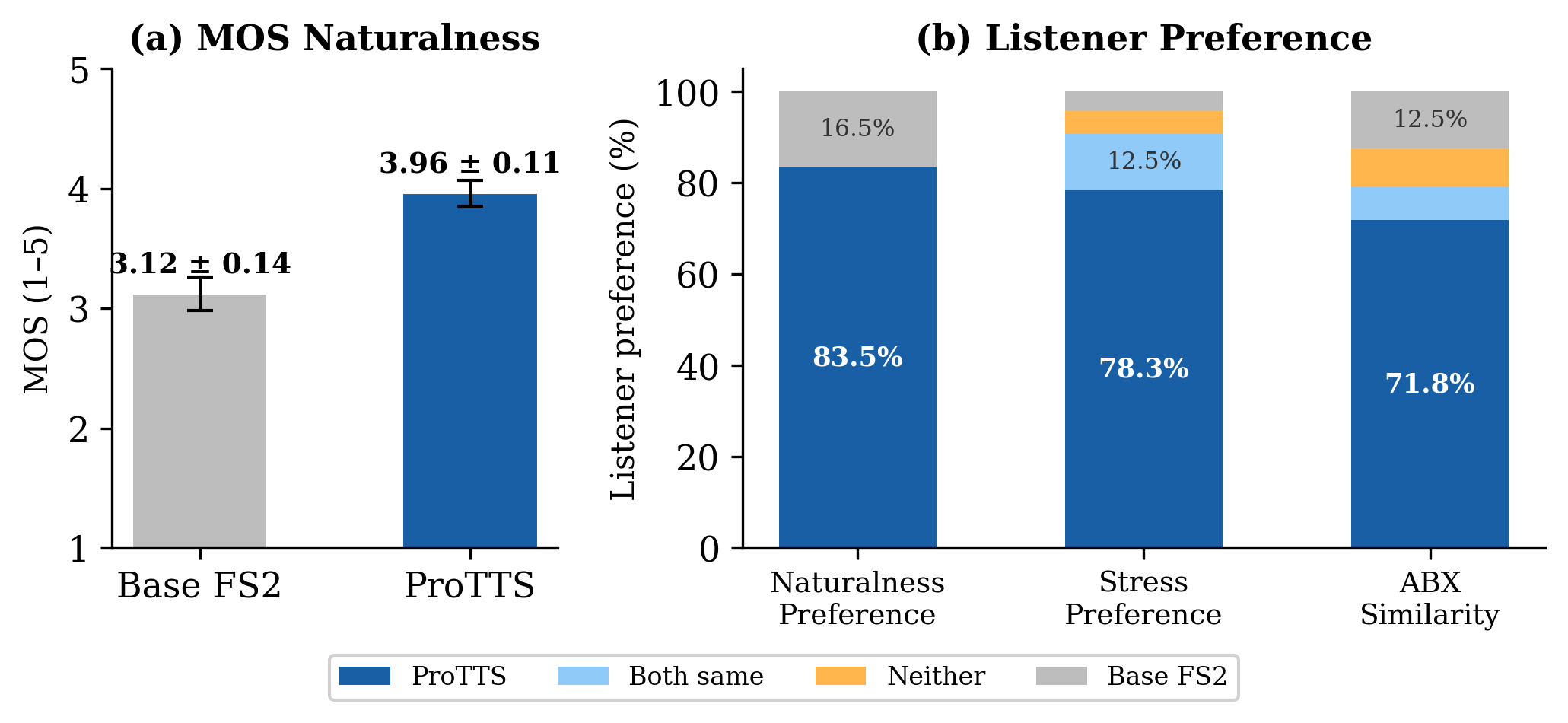}
    \vspace{-0.5cm}
    \caption{Subjective evaluation results. (a) MOS naturalness ratings with 95\% CIs. (b) Listener preference (\%) across naturalness, stress, and ABX similarity tests.}
    \label{fig:synthesis}
    \vspace{-0.2cm}
\end{figure}

\textit{Figure~\ref{fig:synthesis} addresses our third question: whether the discovered stress translates into perceptibly better synthesis.} The listening test was completed by 35 participants with balanced representation across gender, native language, English proficiency, and familiarity with speech and linguistics research (full demographics in Appendix~\ref{app:listening_test}).

The three tests probe different aspects of synthesis quality, evaluating both the automatic and controllable modes of the framework. The \textbf{MOS + preference test} evaluates the automatic mode: the clustering head discovers stress, the emphasis predictors realize it, and listeners judge overall naturalness. ProTTS achieves a MOS of 3.96 $\pm$ 0.11 compared to 3.12 $\pm$ 0.14 for the base FastSpeech\,2 (Figure~\ref{fig:synthesis}a), and is preferred by 83.5\% of listeners (Figure~\ref{fig:synthesis}b). The \textbf{stress preference test} evaluates both modes: a subset of test items uses stress labels predicted by the clustering head (automatic mode), while the remaining items use manually specified stress patterns (controllable mode). Listeners see target stressed words highlighted and judge which audio realizes them better. ProTTS is preferred 78.3\% of the time, with 12.5\% rating both as equal, 5.1\% selecting neither, and only 4.1\% preferring the baseline (Figure~\ref{fig:synthesis}b). This holds across both automatic and controllable modes, confirming that the emphasis variance predictors produce natural and perceptible prosodic prominence whether stress is discovered automatically or specified by the user. Per-dataset results and the automatic vs.\ controllable breakdown are reported in Appendix~\ref{app:synthesis_breakdown}. Finally, the \textbf{ABX similarity test} asks a complementary question: does ProTTS's prosody resemble real speech more closely than the baseline's? This test also includes samples from both automatic and controllable modes. Listeners chose our output as closer to the original recording 71.8\% of the time, compared to 12.5\% for the baseline, with 7.2\% rating both as equal and 8.5\% selecting neither (Figure~\ref{fig:synthesis}b). Together with the stress test, this suggests that the emphasis predictors are not simply exaggerating prosodic contrasts, they are producing stress patterns that align with how speakers naturally realize prominence, regardless of whether the stress mask was discovered or user-specified.

\textbf{Error analysis.} The model handles function words well, words like \emph{the}, \emph{a}, and \emph{of} are rarely misclassified, as their prosodically weak profiles (short duration, reduced vowels, low energy) are clearly captured by the EDP features. The main source of error lies in content words, where two patterns emerge. First, the model tends to over-predict stress on content words that are contextually unstressed. Content words generally carry stronger prosodic profiles (longer duration, higher energy) even when they do not receive sentential stress, and the clustering head sometimes confuses this inherent word-level salience with actual prominence. Second, the model misses stress on monosyllabic content words with short vowels or voiceless consonants (e.g., \emph{pub}, \emph{cup}, \emph{bus}), where the acoustic cues for stress (pitch rise, energy boost, durational lengthening) are inherently weaker and harder to distinguish from unstressed realizations. This difficulty is amplified in non-native speech, where L2 speakers often produce reduced prosodic contrasts between stressed and unstressed words, making the boundary less acoustically salient. These patterns point to a fundamental limitation of relying on acoustic prosodic features alone: stress that depends on discourse context, pragmatic intent, or phonetic properties of specific segments cannot always be recovered from pitch, energy, and duration.

\vspace{-0.1cm}
\subsubsection{Ablation Studies}
\label{sec:ablations}
\begin{table}[t]
\centering
\small
\setlength{\tabcolsep}{4pt}
\begin{tabular}{@{}p{5.0cm}cc@{}}
\toprule
\textbf{Configuration} & \textbf{GER} & \textbf{ITA} \\
\midrule

\multicolumn{3}{@{}l}{\emph{\textbf{(a) Training strategy}}} \\
ProTTS (joint) & 0.839 & 0.861 \\
ProTTS (casc.) & 0.803 & 0.831 \\

\midrule
\multicolumn{3}{@{}l}{\emph{\textbf{(b) Feature composition (centroid init.)}}} \\
Duration + Pitch (DP) & 0.731 & 0.782 \\
Energy + Duration + Pitch (EDP) & 0.664 & 0.675 \\

\midrule
\multicolumn{3}{@{}l}{\emph{\textbf{(c) Native $\rightarrow$ non-native transfer}}} \\
Trained on LibriTTS & 0.781 & 0.752 \\
Trained on VCTK & 0.764 & 0.73 \\

\bottomrule
\end{tabular}
\caption{Ablation results (F1) on ISLE. (a) Joint vs.\ cascaded training. (b) Raw feature composition for K-Means centroid initialization. (c) Native $\rightarrow$ non-native cross-domain transfer.}
\label{tab:ablations}
\vspace{-0.5cm}
\end{table}

\textit{Table~\ref{tab:ablations} addresses our fourth question by examining key design choices.}

\textbf{Joint vs.\ cascaded (Table~\ref{tab:ablations}a).} ProTTS achieves 0.839/0.861 vs.\ 0.803/0.831 for the cascaded baseline (Appendix \ref{app:cascaded}). The improvement confirms that sharing the prosodic feature space between detection and synthesis, rather than passing labels across a pipeline boundary, enables the two tasks to mutually refine each other within a single optimization loop.

\textbf{Feature composition (Table~\ref{tab:ablations}b).} To determine which prosodic features best initialize the clustering centroids, we compared K-Means clustering using two feature subsets from the raw base predictions: duration + pitch (DP) and energy + duration + pitch (EDP). DP yields 0.731/0.782, substantially outperforming EDP at 0.664/0.675. This is consistent with the phonetic literature \citep{vanheuven2018acoustic, gasteiger2024scoping}. Based on this, we use DP features for centroid initialization in all main experiments. Note that this ablation reflects only which raw features initialize centroids, not which features affect the training. During training we use full 521-dim embeddings (EDP: Equation~\ref{eq:train_feats}).

\textbf{Cross-domain transfer (Table~\ref{tab:ablations}c).} 
Training on native English speech and testing on non-native ISLE, the model achieves 0.781/0.752 (LibriTTS) and 0.764/0.73 (VCTK) on GER/ITA, lower than in-domain, as expected, but still reasonable given the domain mismatch. Transfer to German speakers is consistently stronger than to Italian. This reflects typological proximity: German, like English, is stress-timed, so native English stress patterns map more naturally onto German L2 speech, whereas Italian's syllable-timed rhythm diverges more from English prosody.
\section{Conclusion}
\label{sec:conclusion}
\vspace{-0.2cm}
We presented ProTTS, a unified framework that jointly performs unsupervised word stress detection and stress-aware speech synthesis within a single TTS model. By integrating a clustering head into the variance adaptor of FastSpeech\,2, ProTTS discovers stressed words directly from the model's own prosodic predictions (pitch, energy, and duration) without requiring stress annotations at any stage. This stress then guides emphasis variance predictors that modify prosody for stressed words, producing more natural and expressive speech. The same architecture also supports user-controllable emphasis by accepting a custom stress mask at inference. Experiments across native and non-native speech show that ProTTS outperforms both unsupervised and supervised baselines on stress detection, achieves strong zero-shot transfer across corpora and speaker populations, and produces synthesis that listeners consistently prefer over the baseline. The end-to-end objective drives the variance adaptor to learn genuinely stress-discriminative features, an outcome that neither standalone detection nor cascaded approaches achieve. More broadly, our results suggest that TTS models already encode richer prosodic structure than is typically exploited. The prosodic features inside a variance adaptor are not just intermediate predictions for reconstruction, they are a latent space where linguistic phenomena like stress can be discovered and leveraged. We hope this perspective encourages further work on extracting and utilizing the prosodic knowledge embedded within speech synthesis models. Code and trained models will be released upon acceptance.

\section*{Limitations}
\label{sec:limitations}

This work has several limitations that point to directions for future research.

First, as the error analysis reveals, ProTTS relies solely on acoustic prosodic features (pitch, energy, and duration) for stress detection. This works well for function words, whose prosodically weak profiles are clearly captured, but leads to errors on content words where stress depends on discourse context, pragmatic intent, or phonetic properties of specific segments (e.g., short vowels, voiceless consonants). Incorporating text-level features or language model representations could help disambiguate these cases.

Second, the clustering head assumes a binary stressed/unstressed distinction ($K=2$), but stress in natural speech is gradient, words can carry varying degrees of prominence. The current emphasis mechanism applies a binary mask, making it difficult to extend to finer-grained stress levels. Future work could explore soft emphasis scaling proportional to cluster confidence rather than hard assignment.

Third, we evaluate on German and Italian L2 speakers, but stress realization patterns vary across L1 backgrounds. Speakers of tonal languages (e.g., Mandarin, Thai) may exhibit qualitatively different prosodic profiles that interact with English stress in ways not captured by our current evaluation. Extending to typologically diverse L1 backgrounds is important.

Fourth, ProTTS is demonstrated on FastSpeech\,2. While the core idea, integrating a clustering head into the variance adaptor, is architecture-general, we have not evaluated on autoregressive or diffusion-based TTS models, where the variance adaptor structure differs.

Finally, the current design aggregates phoneme-level features into word-level representations by mean-pooling, which dilutes the prosodic signal for polysyllabic words where stress falls on a single syllable. A syllable-level variant of the clustering head could address this and potentially improve performance on longer words.

\section*{Ethical Considerations}
\label{sec:ethics}

This work improves the naturalness and expressiveness of synthesized speech by incorporating word-level stress. While more natural-sounding TTS has broad positive applications, language learning tools, accessibility for visually impaired users, more engaging voice assistants, we acknowledge that improvements in speech synthesis quality could also lower the barrier for generating convincing synthetic speech for malicious purposes (e.g., impersonation or misinformation). We note, however, that ProTTS operates on standard TTS architectures and does not introduce new capabilities for voice cloning or speaker impersonation; the emphasis mechanism modifies prosody within an existing speaker's voice profile. All human evaluation data was collected with informed consent from voluntary participants, and no personally identifiable information was retained. AI writing assistance tools were used during the drafting and editing of this manuscript. All technical content, experimental design, methodology, and reported results are entirely the authors' original work.

\bibliography{custom}

\appendix


\section{Dataset Statistics}
\label{app:datasets}

\begin{figure}[h!]
    \centering
    \includegraphics[width=\columnwidth]{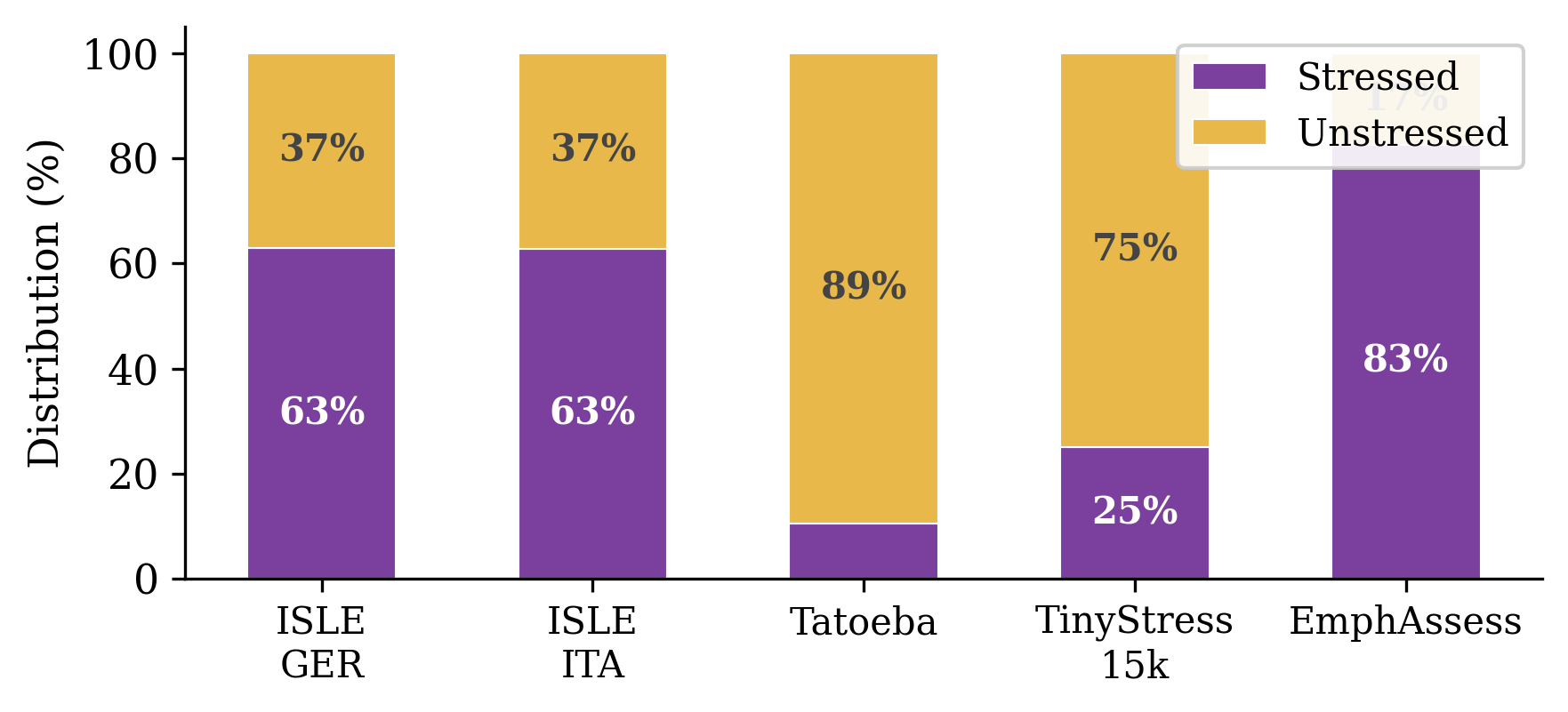}
    \caption{Stressed vs.\ unstressed word distribution across annotated datasets. The substantial variation in stress ratios (10.6\%--82.6\%) tests model robustness across annotation conventions.}
    \label{fig:stress_dist}
\caption{Stress label distribution across annotated datasets.}
\label{tab:stress_dist}
\end{figure}

Table~\ref{tab:datasets} summarizes all datasets used for training and evaluation. Table~\ref{tab:stress_dist} reports the word-level stress label distribution for all annotated datasets.

\begin{table}[h!]
\centering
\small
\setlength{\tabcolsep}{3pt}
\resizebox{\columnwidth}{!}{
\begin{tabular}{@{}lrrrcr@{}}
\toprule
\textbf{Dataset} & \textbf{Train} & \textbf{Val} & \textbf{Test} & \textbf{Spkrs} & \textbf{Duration} \\
\midrule
ISLE (GER+ITA) & 3,750 & 3,252 & --- & 46 & 5.09 h \\
LibriTTS (clean-100) & 3,000 & 469 & --- & 264 & 59.23 h \\
VCTK & 3,000 & 509 & --- & 20 & 4.35 h \\
\midrule
Tatoeba & 3,000 & 398 & 993 & diverse & --- \\
TinyStress-15k & --- & --- & 1,000 & synth. & --- \\
EmphAssess & --- & --- & 3,652 & 4 & 2.08 h \\
\bottomrule
\end{tabular}
}
\caption{Dataset statistics. Splits reported in utterances. Tatoeba is a community-contributed spoken English corpus \citep{tatoeba}. TinyStress-15k is a synthetic dataset based on TinyStories \citep{yosha2025whistress}; we use only the test split. EmphAssess \citep{deseyssel2024emphassess} is a prosodic benchmark with emphasis annotations. ISLE stress evaluation uses the validation split.}
\label{tab:datasets}
\end{table}

The stress distributions vary substantially across datasets. ISLE exhibits a relatively balanced distribution ($\sim$63\% stressed), reflecting the read-aloud nature of the corpus where most content words receive prominence. Tatoeba is heavily skewed toward unstressed words (10.6\% stressed), as the annotations mark only sentential stress on key content words in conversational sentences. EmphAssess shows the opposite extreme (82.6\% stressed), consistent with its design as an emphasis evaluation benchmark where speakers are instructed to produce prominent realizations. TinyStress-15k falls between these at 25\% stressed. This diversity in label distributions tests the robustness of ProTTS across different stress definitions and annotation conventions.
\begin{table}[h!]
\centering
\small
\begin{tabular}{@{}lrrrr@{}}
\toprule
\textbf{Dataset} & \textbf{Words} & \textbf{Stressed} & \textbf{Unstressed} & \textbf{\% Str.} \\
\midrule
ISLE GER & 27,136 & 17,082 & 10,054 & 62.9 \\
ISLE ITA & 25,225 & 15,848 & 9,377 & 62.8 \\
Tatoeba & 6,720 & 712 & 6,008 & 10.6 \\
TinyStress-15k & 195,000 & 48,752 & 146,248 & 25.0 \\
EmphAssess & 35,371 & 29,216 & 6,155 & 82.6 \\
\bottomrule
\end{tabular}
\end{table}

\section{Stress Annotation Procedure}
\label{app:annotation}

\textbf{Tatoeba.} We selected a subset of English audio files with transcripts from the Tatoeba platform \citep{tatoeba}, a community-driven repository of multilingual sentences and recordings. Each utterance was manually annotated at the word level for binary prominence (1 = stressed, 0 = unstressed), excluding silences and noise segments. The annotated subset contains 7,122 words, of which 1,083 are labeled as stressed and 6,039 as unstressed. We partitioned this into 3,000 training, 398 validation, and 993 test utterances; only the test split is used for evaluation.

\textbf{ISLE corpus.} The ISLE corpus \citep{menzel2000isle} provides speech from 46 non-native English speakers (23 German, 23 Italian) with phoneme-level transcriptions and syllable-level prominence annotations (48,868 stressed and 16,693 unstressed syllables in the full corpus). Since ProTTS operates at the word level, we additionally annotated a subset of 2,000 utterances (1,000 per L1 background) with binary word-level stress labels, following the same annotation protocol as Tatoeba. Word boundaries were derived from the existing phoneme alignments. Training and validation splits were constructed to maintain balanced representation across speaker attributes including nativity, age, gender, and proficiency level.
\begin{table*}[t]
\centering
\small
\setlength{\tabcolsep}{6pt}
\renewcommand{\arraystretch}{1.15}

\resizebox{0.82\textwidth}{!}{%
\begin{tabular}{p{2.8cm}p{10.5cm}}
\toprule
\textbf{Symbol} & \textbf{Definition} \\
\midrule

$\mathbf{x} = (x_1,\ldots,x_N)$
& Input phoneme sequence. \\

$\mathbf{h} = (h_1,\ldots,h_N)$
& Encoder hidden representations corresponding to the phoneme sequence. \\

$w_k$
& $k$-th word spanning phonemes $x_{s_k}\ldots x_{e_k}$. \\

$\hat{p}_i,\ \hat{e}_i,\ \hat{d}_i$
& Predicted pitch, energy, and duration from the variance adaptor. \\

$\mathbf{p}_i^{\text{emb}},\ \mathbf{e}_i^{\text{emb}}$
& Pitch and energy embeddings associated with phoneme $x_i$. \\

$\mathbf{z}_k$
& Aggregated word-level feature representation used for clustering. \\

$\mathbf{z}_k^{(p)},\ \mathbf{z}_k^{(e)},\ \mathbf{z}_k^{(d)}$
& Pitch, energy, and duration components of $\mathbf{z}_k$. \\

$\boldsymbol{\mu}_0,\ \boldsymbol{\mu}_1$
& Learnable clustering centroids corresponding to unstressed and stressed groups. \\

$q_{kj}$
& Soft assignment probability of word $k$ belonging to cluster $j$. \\

$p_{kj}$
& Sharpened target distribution derived from $q_{kj}$. \\

$f_j$
& Soft frequency of cluster $j$ used for balancing. \\

$y_k$
& Hard pseudo-label assigned to word $k$. \\

$m_k \in \{0,1\}$
& Emphasis mask predicted from the clustering head (0 = unstressed, 1 = stressed). \\

$\Delta \hat{p}_i,\ \Delta \hat{e}_i,\ \Delta \hat{d}_i$
& Predicted emphasis-induced prosodic deltas. \\

$\tilde{p}_i,\ \tilde{e}_i,\ \tilde{d}_i$
& Final stress-adapted pitch, energy, and duration values. \\

$\mathcal{L}_{\text{recon}}$
& Reconstruction loss for mel-spectrogram and variance prediction. \\

$\mathcal{L}_{\text{kl}}$
& KL-divergence clustering objective. \\

$\mathcal{L}_{\text{bal}}$
& Cluster balancing regularization loss. \\

$\mathcal{L}_{\text{total}}$
& Overall training objective combining reconstruction and clustering losses. \\

$\alpha$
& Weight controlling the contribution of clustering-related losses. \\

\bottomrule
\end{tabular}%
}

\caption{Notation used throughout the paper.}
\label{tab:notation}
\end{table*}
\textbf{TinyStress-15k.} This synthetic dataset \citep{yosha2025whistress} contains approximately 15,000 utterances (roughly 15 hours) generated from TinyStories text. Stress labels were produced automatically by prompting GPT-4o-mini to identify semantically important words, and the corresponding speech was synthesized using the Google TTS API with SSML markup to explicitly modulate amplitude, pitch, and speaking rate for stressed words. We use only the 1,000-utterance test split for evaluation.

\textbf{EmphAssess.} This benchmark \citep{deseyssel2024emphassess} contains 3,652 utterances derived from 913 unique transcripts, synthesized using four Expresso TTS voices \citep{nguyen2023expresso}. Each utterance is designed to contain at least one emphasized word, with emphasis realized through controlled voice rendering parameters rather than post-hoc signal manipulation. The use of multiple synthetic voices provides speaker variability while ensuring consistent emphasis placement.

\section{Notation}
\label{app:notation}

For convenience, Table~\ref{tab:notation} summarizes the key symbols and notation used throughout the paper, organized by the component they belong to.


\section{Hyperparameters and Training Details}
\label{app:hyperparams}

The model is initialized from a FastSpeech\,2 checkpoint pretrained for 900k steps. We use the Adam optimizer with an initial learning rate of $1 \times 10^{-5}$ and a cosine annealing scheduler with warm restarts, batch size of 8, and train for 100,000 steps. The clustering weight is $\alpha = 1.0$. The emphasis variance predictors follow the same architecture as the base variance predictors (2-layer Conv1D, kernel size 3, ReLU, dropout 0.5), with output layers zero-initialized. K-Means initialization uses 20 random restarts on the duration + pitch feature subset. The HiFi-GAN vocoder is pretrained on LJSpeech.

All experiments are conducted on a single NVIDIA GPU. The base FastSpeech\,2 model contains approximately 30M parameters. The three emphasis variance predictors add $\sim$1.5M parameters, and the clustering head adds minimal overhead (2 centroids $\times$ 521 dimensions $\approx$ 1K parameters), bringing the total to approximately 31.5M parameters. Inference overhead is negligible since the cluster assignment is a single distance computation per word.

\section{Word-Level Feature Components}
\label{app:features}

The word-level feature vector $\mathbf{z}_k \in \mathbb{R}^{521}$ (Section~\ref{sec:aggregation}) is constructed by concatenating three components. For word $w_k$ spanning phonemes $\{x_{s_k}, \ldots, x_{e_k}\}$:

\textbf{Pitch embedding} ($\mathbb{R}^{256}$): Mean-pooled phoneme-level pitch embeddings:
\begin{equation}
    \mathbf{z}_k^{(p)} = \frac{1}{e_k - s_k + 1} \sum_{i=s_k}^{e_k} \mathbf{p}_i^{\text{emb}}
\end{equation}

\textbf{Energy embedding} ($\mathbb{R}^{256}$): Mean-pooled phoneme-level energy embeddings:
\begin{equation}
    \mathbf{z}_k^{(e)} = \frac{1}{e_k - s_k + 1} \sum_{i=s_k}^{e_k} \mathbf{e}_i^{\text{emb}}
\end{equation}

\textbf{Duration and positional statistics} ($\mathbb{R}^{9}$):
\begin{equation}
\begin{split}
    \mathbf{z}_k^{(d)} = \big[&\text{mean}(\hat{d}_{s_k:e_k}),\; \text{std}(\hat{d}_{s_k:e_k}),\; \max(\hat{d}_{s_k:e_k}), \\
    &\min(\hat{d}_{s_k:e_k}),\; \text{sum},\; \text{range},\; \text{cv},\; \text{pos}_k,\; n_k^{\text{ph}} \big]
\end{split}
\end{equation}

\section{Training Procedure Details}
\label{app:training}

The model is initialized from a pretrained FastSpeech\,2 checkpoint. At epoch 0, the clustering centroids are initialized from the base model's prosodic features (Section~\ref{sec:clustering}) and the emphasis predictor output layers are zero-initialized so that the stress-adapted prosody is initially identical to the base prosody. Each training epoch proceeds as follows:

\begin{enumerate}
    \item \textbf{Forward pass:} The encoder produces hidden states; the base variance predictors generate the base prosody ($\hat{p}_i$, $\hat{e}_i$, $\hat{d}_i$); these are aggregated into 521-dim word features; the clustering head computes soft assignments; the emphasis mask gates the emphasis predictors to produce the stress-adapted prosody ($\tilde{p}_i$, $\tilde{e}_i$, $\tilde{d}_i$); the stress-adapted prosody passes through the length regulator and decoder to produce the mel-spectrogram.
    \item \textbf{Backward pass:} A single backward pass through $\mathcal{L}_{\text{total}}$ updates all model parameters jointly, including the encoder, base predictors, clustering centroids, emphasis predictors, and decoder.
    \item \textbf{Pseudo-label update:} Word-level embeddings are collected for all training utterances and hard assignments (Eq.~\ref{eq:hard_assign}) are computed to update the emphasis pseudo-labels.
    \item \textbf{Variance target regeneration:} The updated pseudo-labels determine which words are considered stressed, and variance targets are recomputed accordingly.
\end{enumerate}

\section{Detailed Loss Formulation}
\label{app:losses}

The full set of prosody prediction losses in $\mathcal{L}_{\text{recon}}$ (Section~\ref{sec:training}):

\textbf{Combined prosody losses} (stress-adapted prosody vs.\ ground truth):
\begin{align}
    \mathcal{L}_{\text{pitch}} &= \text{MSE}(\hat{p}_i + \Delta \hat{p}_i,\; p_i^{\text{gt}}) \\
    \mathcal{L}_{\text{energy}} &= \text{MSE}(\hat{e}_i + \Delta \hat{e}_i,\; e_i^{\text{gt}}) \\
    \mathcal{L}_{\text{dur}} &= \text{MSE}(\hat{d}_i + \Delta \hat{d}_i,\; d_i^{\text{gt}})
\end{align}

\textbf{Variance losses} (emphasis deltas vs.\ variance targets):
\begin{align}
    \mathcal{L}_{\text{pv}} &= \text{MSE}(\Delta \hat{p}_i,\; \Delta p^{\text{gt}}) \\
    \mathcal{L}_{\text{dv}} &= \text{MSE}(\Delta \hat{d}_i,\; \Delta d^{\text{gt}}) \\
    \mathcal{L}_{\text{ev}} &= 0.5 \cdot \text{MSE}(\Delta \hat{e}_i,\; \Delta e^{\text{gt}})
\end{align}
where the variance targets $\Delta p^{\text{gt}}$, $\Delta d^{\text{gt}}$, $\Delta e^{\text{gt}}$ are applied uniformly to all phonemes within the emphasized region; non-emphasized phonemes receive zero targets.

\textbf{Variance targets} follow and extend the formulation of \citet{li2025emetss}. For each utterance, the variance features capture the difference between the average prosody of emphasized phonemes and the sentence-level average:
\begin{align}
    \Delta p^{\text{gt}} &= \bar{p}^{\text{emph}} - \bar{p}^{\text{sent}} \nonumber \\
    \Delta d^{\text{gt}} &= \bar{d}^{\text{emph}} - \bar{d}^{\text{sent}} \\
    \Delta e^{\text{gt}} &= \bar{e}^{\text{emph}} - \bar{e}^{\text{sent}} \nonumber
\end{align}
where $\bar{p}^{\text{emph}}$, $\bar{d}^{\text{emph}}$, and $\bar{e}^{\text{emph}}$ are the mean pitch, duration, and energy over phonemes in emphasized regions (as determined by the current pseudo-labels), and $\bar{p}^{\text{sent}}$, $\bar{d}^{\text{sent}}$, $\bar{e}^{\text{sent}}$ are the corresponding sentence-level means. While \citet{li2025emetss} model only pitch and duration variance, we extend this to include energy, as our ablation on feature composition (Section~\ref{sec:ablations}) shows that energy contributes to stress discriminability after joint training. The energy variance loss is down-weighted by 0.5 to account for its lesser impact. Variance features are normalized to $[0, 2]$ based on data distribution for stability, and are recomputed each epoch as pseudo-labels evolve.

\section{Cascaded Baseline Architecture}
\label{app:cascaded}

\begin{figure}[t]
    \centering
    \includegraphics[width=\columnwidth]{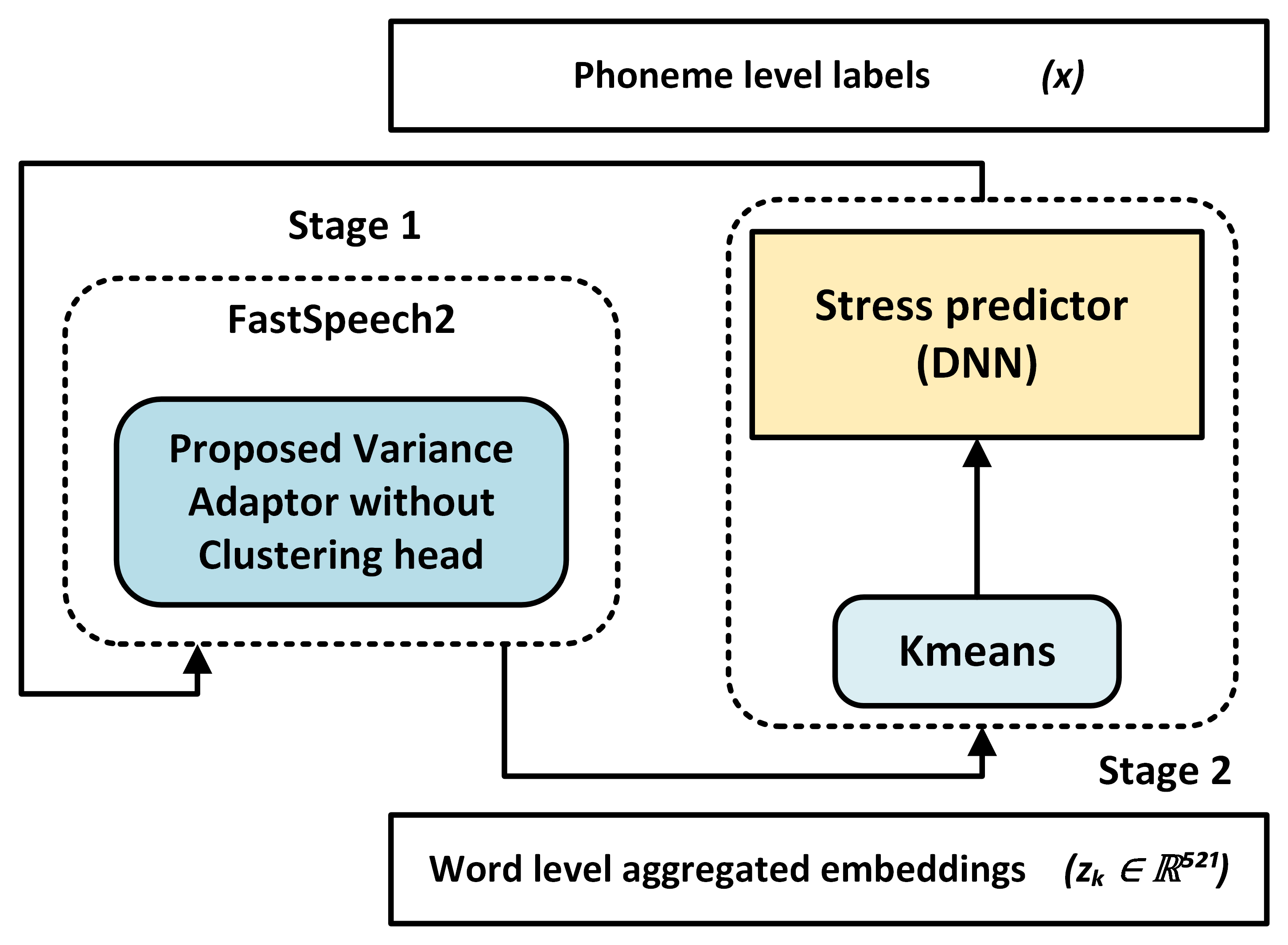}
    \caption{Cascaded (2-stage) baseline architecture. Stage 1 trains FastSpeech\,2 with ProTTS's variance adaptor but without the clustering head. Stage 2 applies K-Means to the word-level aggregated embeddings $\mathbf{z}_k$, then trains a DNN stress predictor on the resulting labels. The two stages alternate but are never trained jointly --- dashed borders indicate separate optimization.}
    \label{fig:cascaded}
\end{figure}

Figure~\ref{fig:cascaded} illustrates the cascaded baseline used in our ablation study (Section~\ref{sec:ablations}). Unlike the joint framework, the cascaded approach separates stress detection and synthesis into two independently optimized stages that alternate iteratively.

\textbf{Stage 1 (FastSpeech\,2 training).} The model uses our proposed variance adaptor with base and emphasis variance predictors, but without the clustering head. In Iteration 0, all words are treated as unstressed ($m_k = 0$), so only the base predictors are active and the emphasis predictors produce zero deltas. From Iteration 1 onward, pseudo-labels from Stage 2 serve as the emphasis mask, activating the emphasis predictors for words labeled as stressed.

\textbf{Stage 2 (Stress classification).} After Stage 1 converges, the word-level aggregated embeddings $\mathbf{z}_k \in \mathbb{R}^{521}$ are extracted from the trained model. K-Means clustering is applied to these embeddings to produce initial binary stress labels. A DNN classifier is then trained on these labels. At inference within the pipeline, the DNN's predictions are used when its confidence exceeds 0.8; otherwise, the K-Means assignment is retained as a fallback.

\textbf{Iterative alternation.} The two stages repeat: updated pseudo-labels from Stage 2 feed back into Stage 1 for the next round of FS2 fine-tuning, and the resulting embeddings are re-clustered and re-classified in Stage 2. Each stage improves from the other's outputs, but they are never co-optimized --- gradients do not flow across the stage boundary. This is the key architectural difference from ProTTS, where the clustering head and emphasis predictors share the same forward-backward pass and mutually shape the prosodic feature space.


\section{Training Convergence}
\label{app:convergence}

\begin{figure}[t]
    \centering
    \includegraphics[width=\columnwidth]{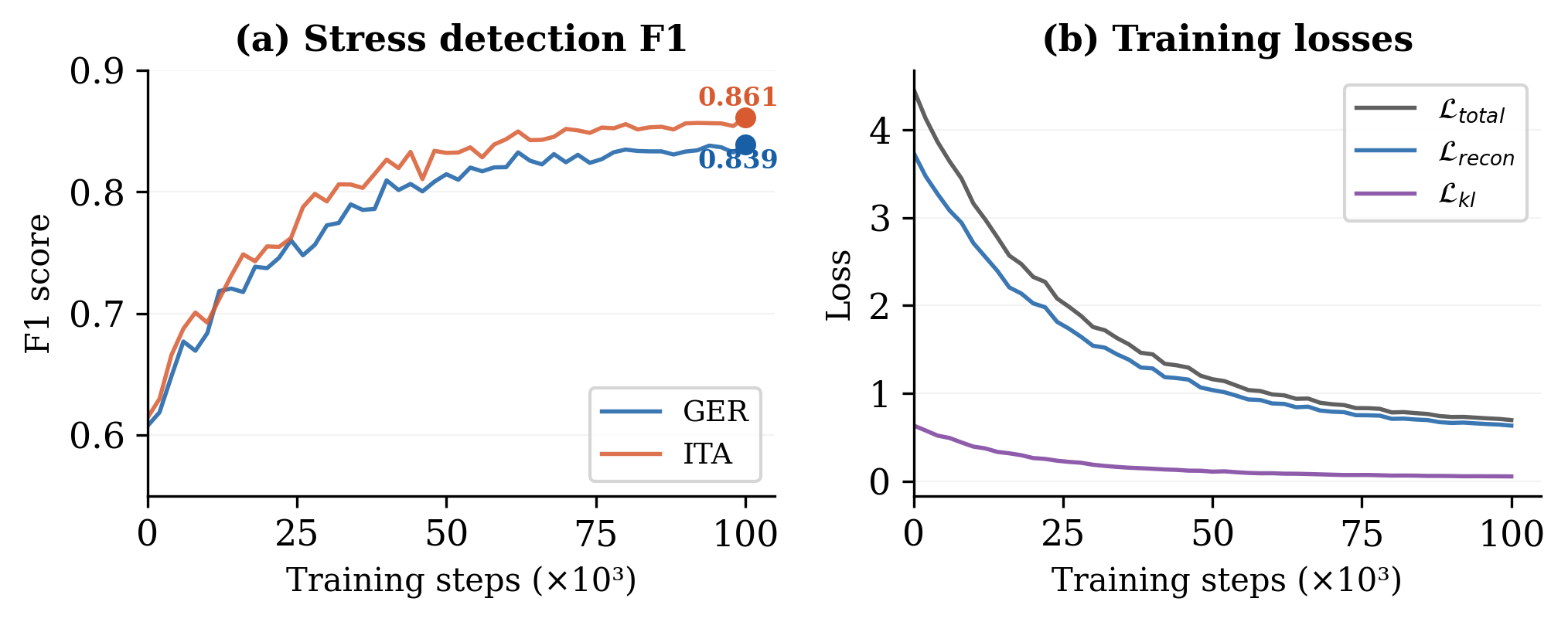}
    \caption{Training convergence over 100k steps. (a) Word stress detection F1 on ISLE validation set for GER and ITA speakers, evaluated at 5k-step intervals. (b) Training losses: total loss $\mathcal{L}_{\text{total}}$, reconstruction loss $\mathcal{L}_{\text{recon}}$, and clustering KL loss $\mathcal{L}_{\text{kl}}$.}
    \label{fig:convergence}
\end{figure}

Figure~\ref{fig:convergence} shows the convergence behavior of the joint training over 100k steps. Stress detection F1 (Figure~\ref{fig:convergence}a) rises rapidly in the first 20--30k steps as the clustering head begins to separate stressed from unstressed words, then gradually plateaus. ITA consistently achieves slightly higher F1 than GER throughout training, consistent with the final results reported in Table~\ref{tab:stress_nonnative}. The training losses (Figure~\ref{fig:convergence}b) decrease smoothly, with the KL clustering loss converging faster than the reconstruction loss, indicating that the clustering head stabilizes before the synthesis pathway has fully adapted. No instability or oscillation is observed, confirming that the joint objective produces stable co-training dynamics.


\section{Listening Test Interface and Instructions}
\label{app:listening_test}

\subsection{Study Design}

\begin{figure}[t]
    \centering
    \includegraphics[width=\columnwidth]{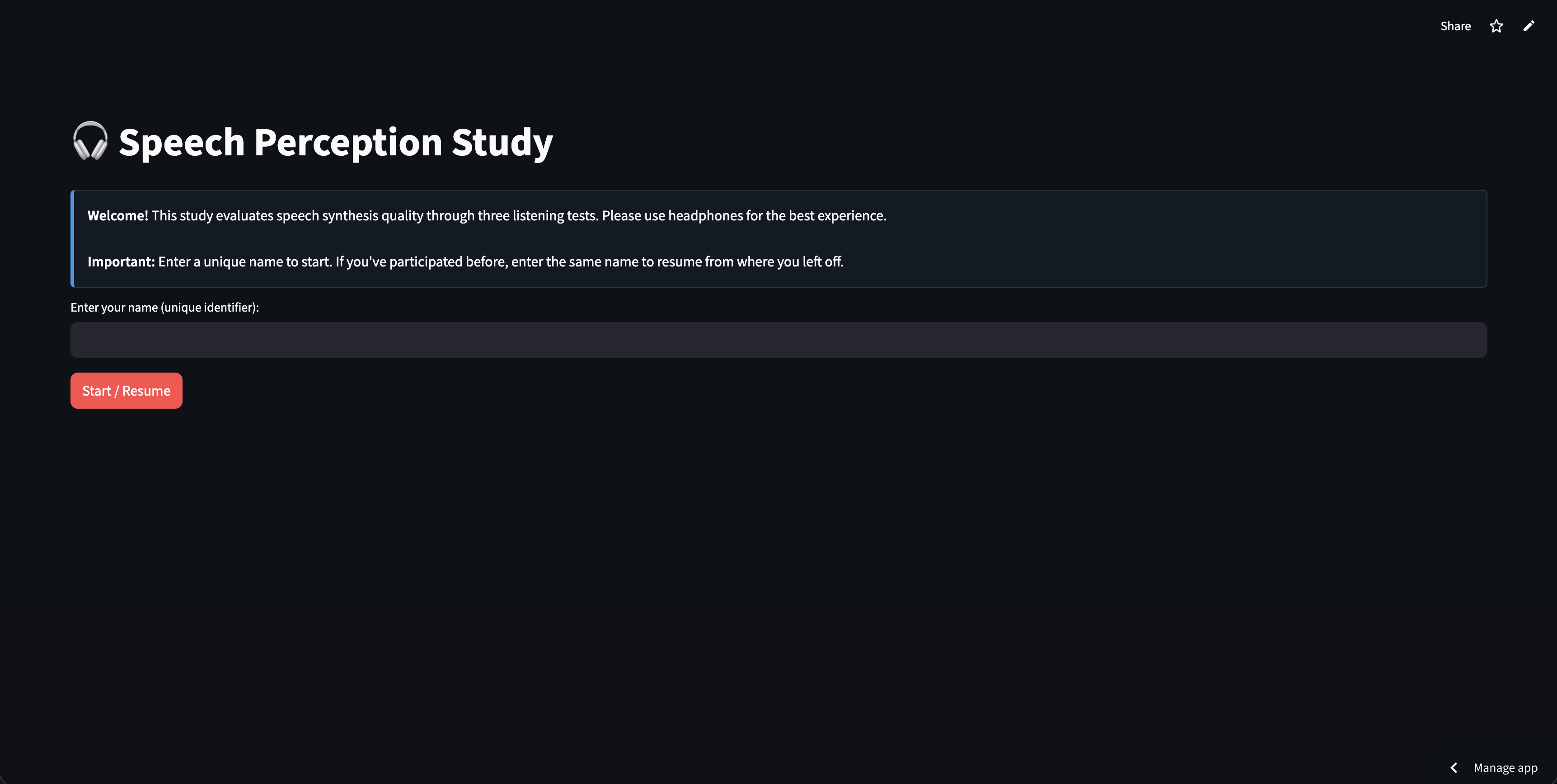}
    \caption{Welcome page of the listening test platform.}
    \label{fig:welcome}
\end{figure}

\begin{figure}[t]
    \centering
    \includegraphics[width=\columnwidth]{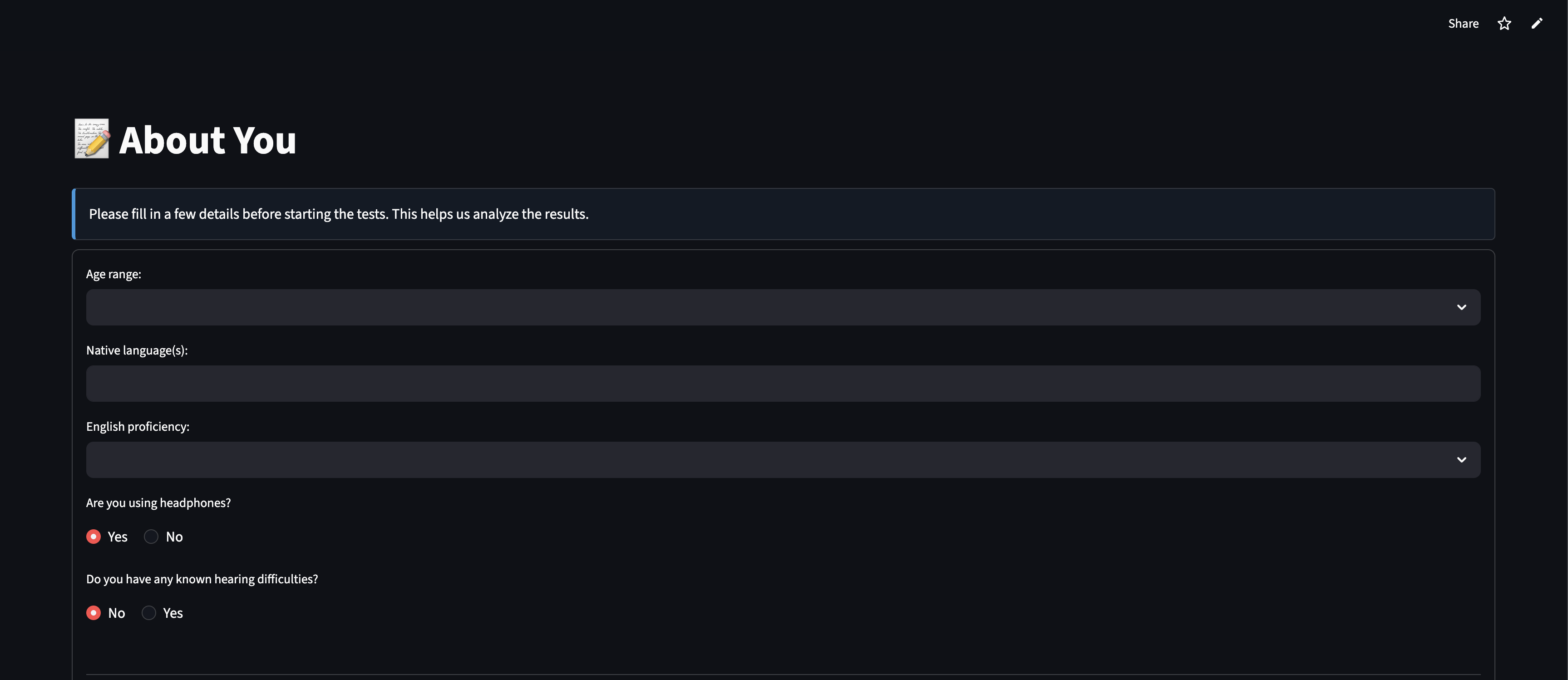}
    \caption{Demographic questionnaire collecting age range, native language, English proficiency, headphone use, and hearing status.}
    \label{fig:demographics}
\end{figure}

\begin{figure}[t]
    \centering
    \includegraphics[width=\columnwidth]{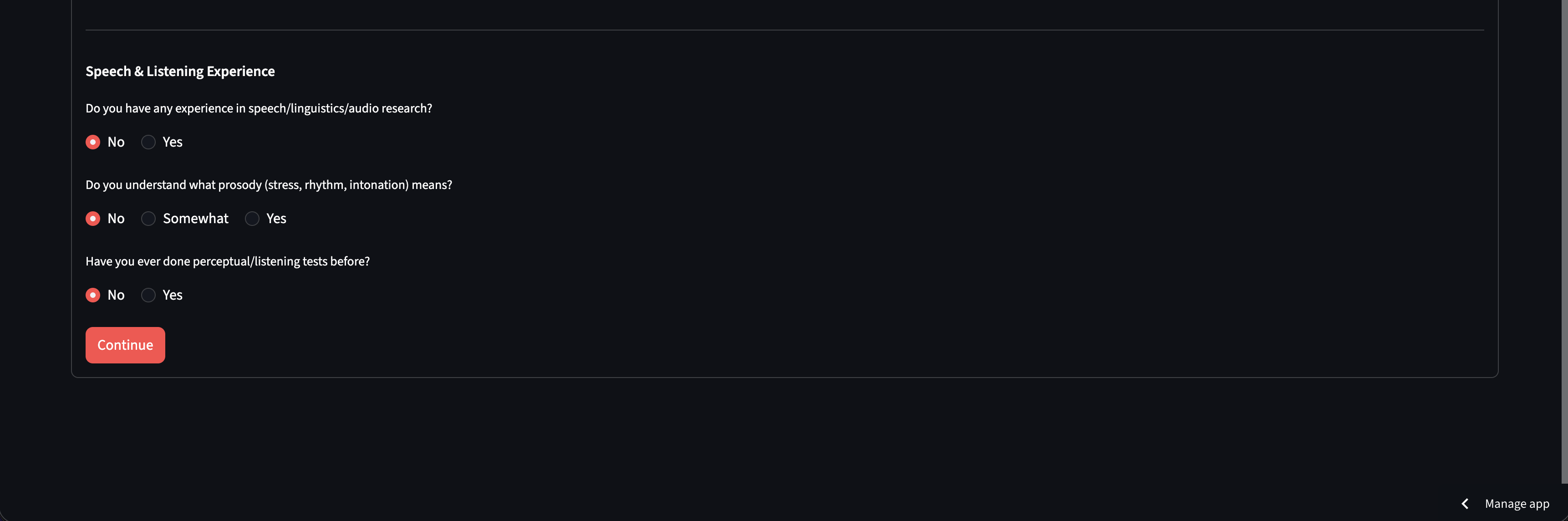}
    \caption{Speech and listening experience survey assessing background in speech/linguistics research, understanding of prosody, and prior perceptual test experience.}
    \label{fig:experience}
\end{figure}

\begin{figure}[t]
    \centering
    \includegraphics[width=\columnwidth]{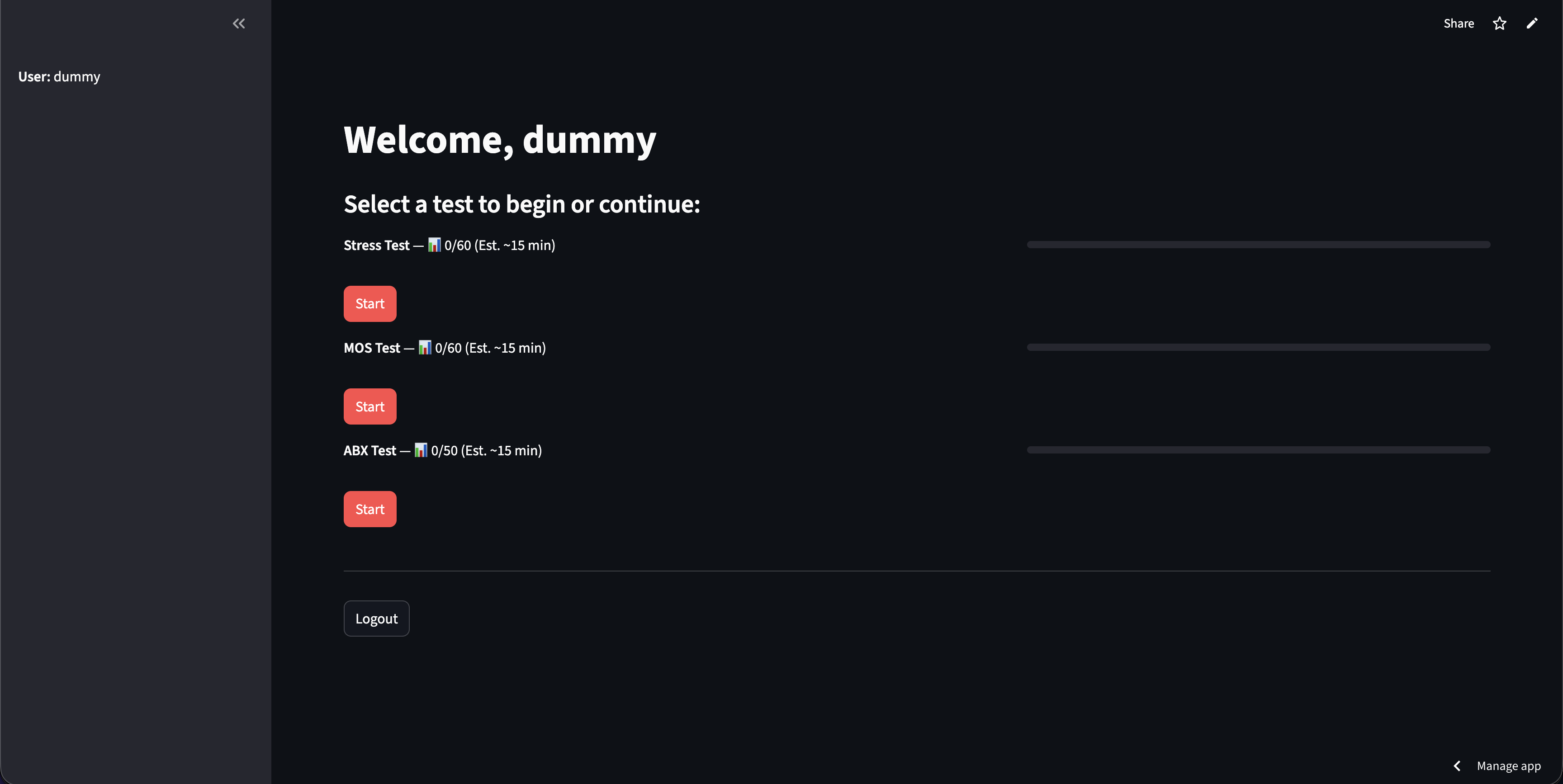}
    \caption{Test selection page. Participants complete three tests: Stress Test (60 questions, $\sim$15 min), MOS Test (60 questions, $\sim$15 min), and ABX Test (50 questions, $\sim$15 min).}
    \label{fig:test_selection}
\end{figure}

We conducted a subjective evaluation study comprising three listening tests administered through a custom web-based platform. Participants accessed the study via a unique identifier and completed all three tests in a single session. The study flow consisted of: (1) a welcome page explaining the study purpose and recommending headphone use (Figure~\ref{fig:welcome}), (2) a demographic questionnaire (Figure~\ref{fig:demographics}), (3) a speech and listening experience survey (Figure~\ref{fig:experience}), and (4) the three listening tests in sequence (Figure~\ref{fig:test_selection}).

\subsection{Participant Demographics}

A total of 35 participants completed all three listening tests. Table~\ref{tab:demographics} summarizes the demographic distribution. Participants were recruited to ensure balanced representation across gender, native language background, English proficiency, and familiarity with speech research.

\begin{table}[t]
\centering
\small
\begin{tabular}{@{}ll@{}}
\toprule
\textbf{Attribute} & \textbf{Distribution} \\
\midrule
Total participants & 35 \\
Gender & Balanced (approx.\ 50/50) \\
Age range & 20--35 \\
Native language & Mix of English and non-English L1 \\
English proficiency & Advanced to native \\
Headphone use & $>$90\% \\
Speech/linguistics background & $\sim$50\% \\
Prior listening test experience & $\sim$45\% \\
Hearing difficulties & None reported \\
\bottomrule
\end{tabular}
\caption{Participant demographics.}
\label{tab:demographics}
\end{table}

The mix of expert and non-expert listeners allows us to assess whether the perceived stress differences are robust across listener backgrounds, rather than being detectable only by trained phoneticians.

\subsection{Test Details}

\begin{figure}[t]
    \centering
    \includegraphics[width=\columnwidth]{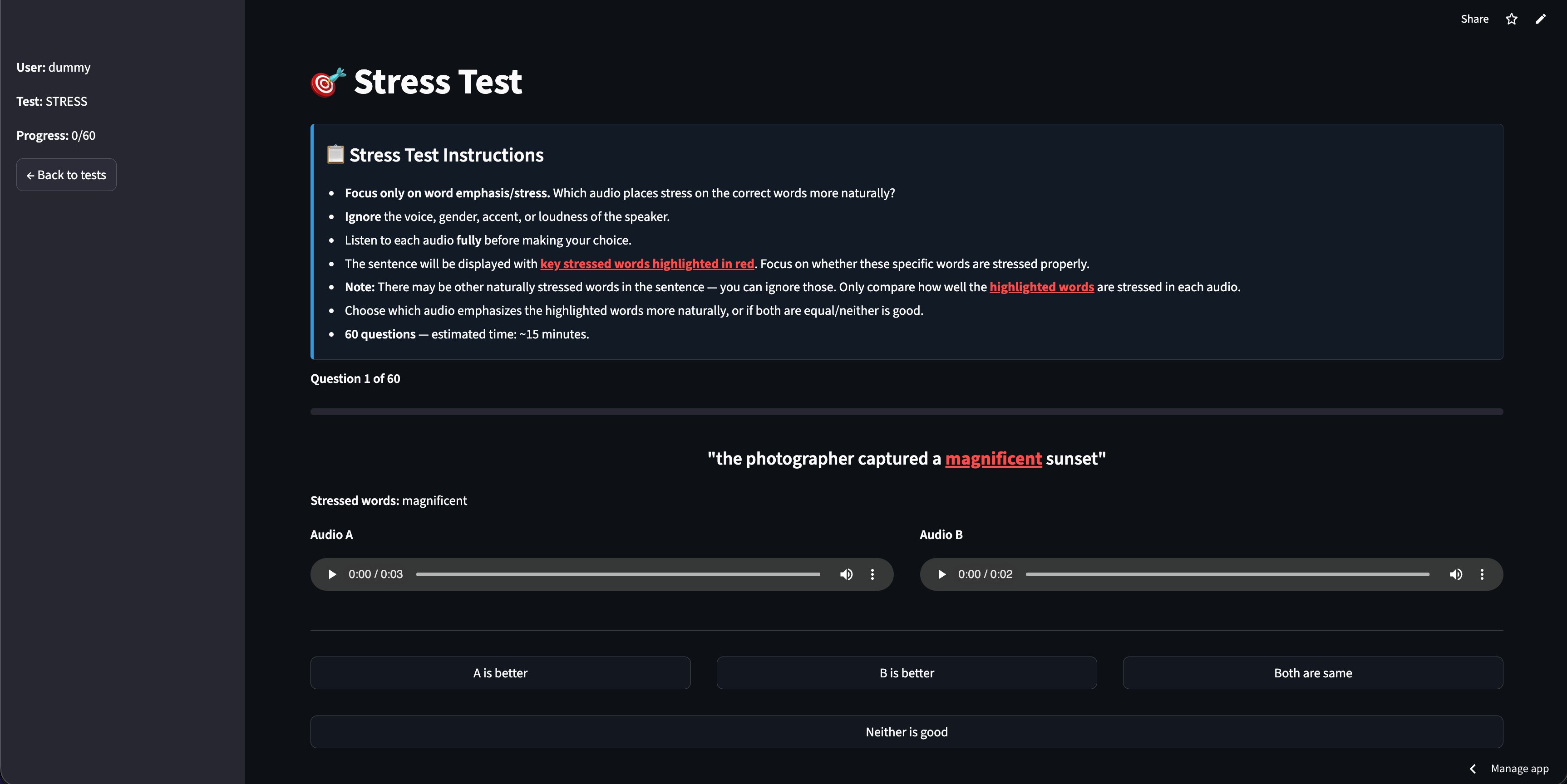}
    \caption{Stress preference test interface.}
    \label{fig:stress_interface}
\end{figure}

\textbf{Stress preference test} (60 questions, $\sim$15 min). The target sentence is displayed with stressed words highlighted in red. Listeners hear two audio samples (randomized order) and select which better realizes the intended stress, with four options: ``A is better,'' ``B is better,'' ``Both are same,'' and ``Neither is good.'' Figure~\ref{fig:stress_interface} shows the interface.

\begin{figure}[t]
    \centering
    \includegraphics[width=\columnwidth]{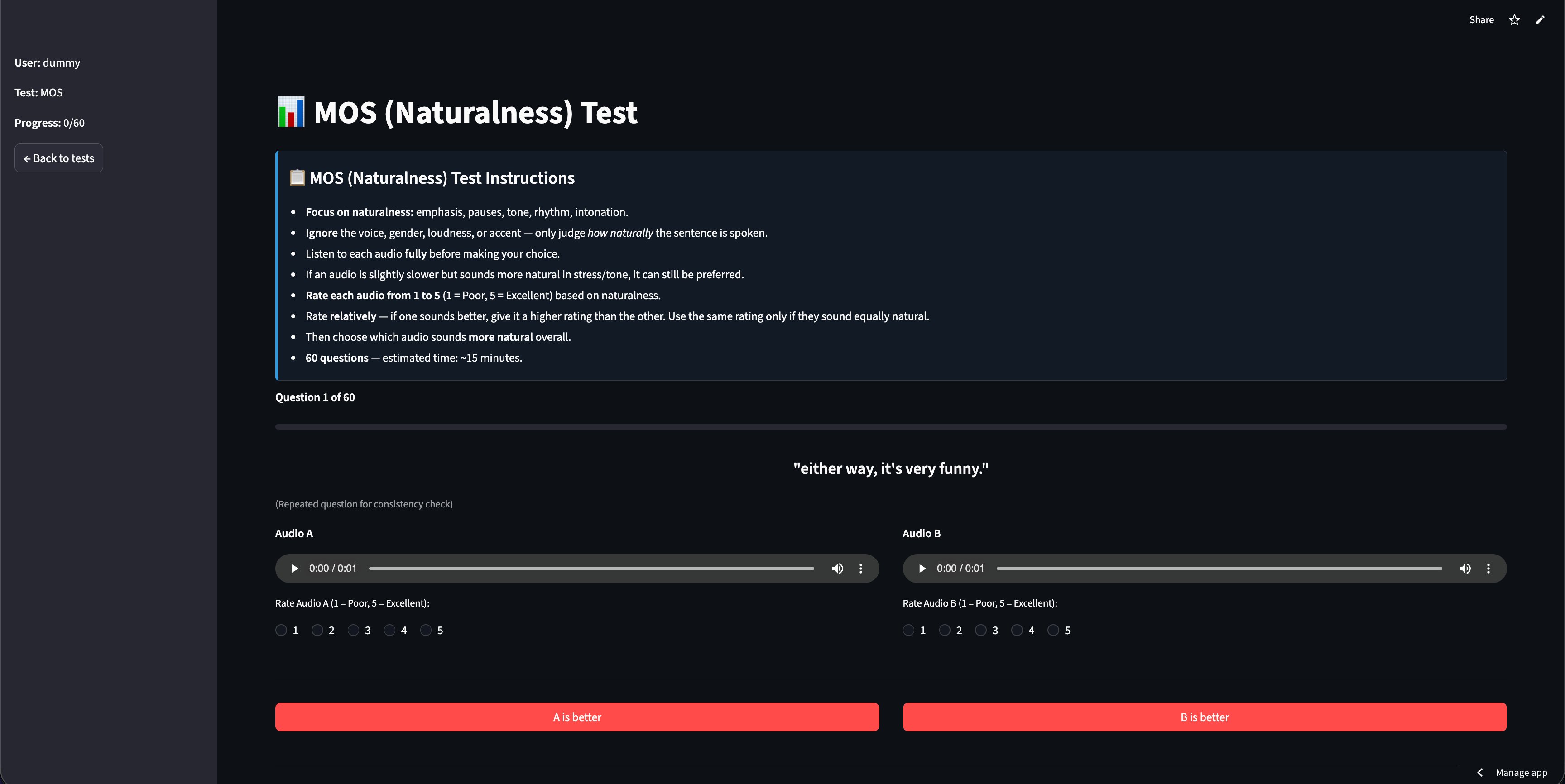}
    \caption{MOS (Naturalness) test interface.}
    \label{fig:mos_interface}
\end{figure}

\textbf{MOS + Preference test} (60 questions, $\sim$15 min). Listeners hear two audio samples for the same sentence, rate each independently on a 1--5 scale (1 = Poor, 5 = Excellent) for naturalness, and then select which sounds more natural overall. Instructions emphasize focusing on emphasis, pauses, tone, and rhythm while ignoring voice, gender, loudness, or accent. Figure~\ref{fig:mos_interface} shows the interface.

\begin{figure}[t]
    \centering
    \includegraphics[width=\columnwidth]{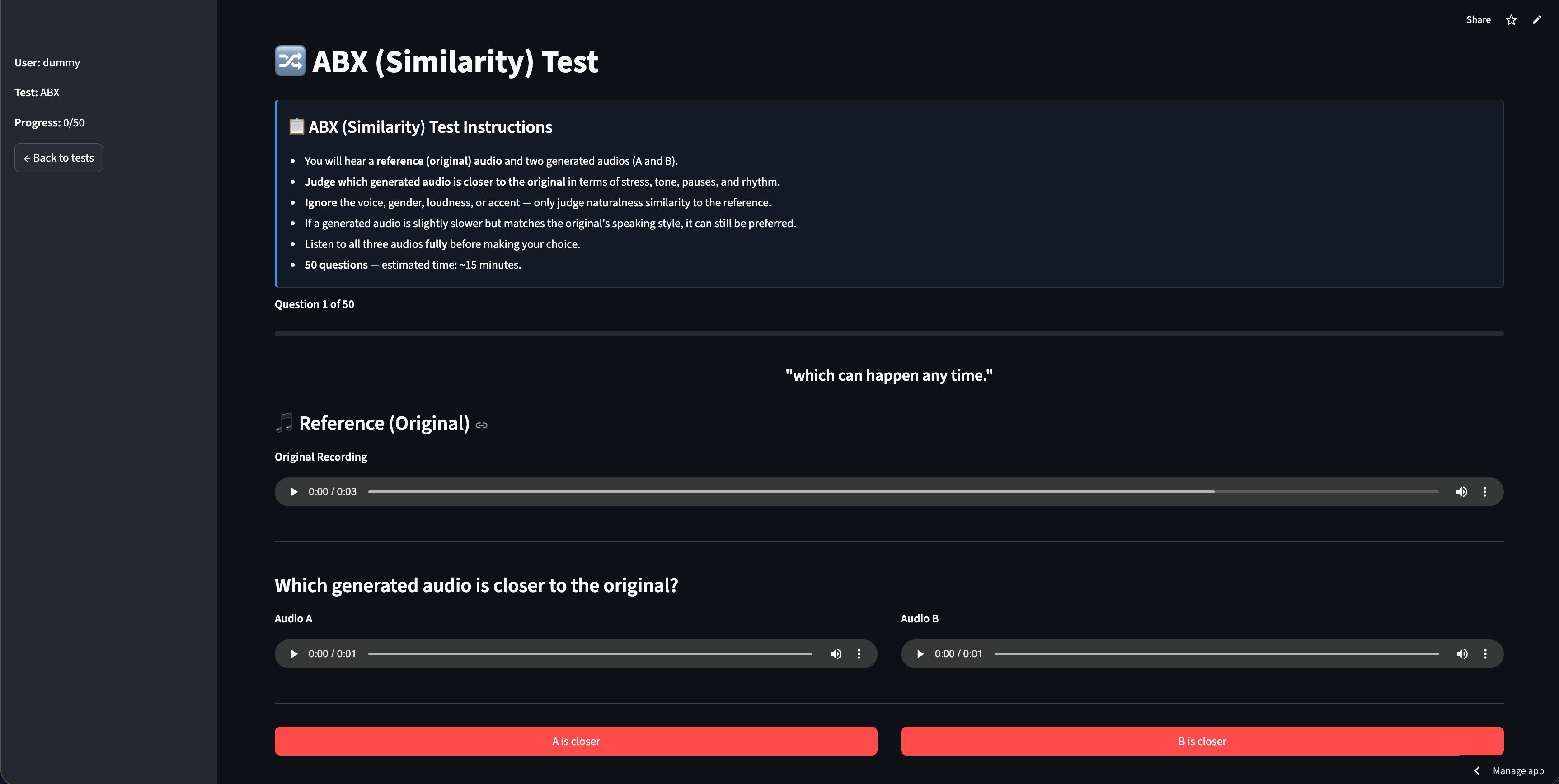}
    \caption{ABX (Similarity) test interface.}
    \label{fig:abx_interface}
\end{figure}

\textbf{ABX similarity test} (50 questions, $\sim$15 min). Listeners hear a reference (original) recording and two generated audio samples (A and B), and judge which generated audio is closer to the original in terms of stress, tone, pauses, and rhythm. Instructions specify ignoring voice, gender, loudness, or accent differences. Figure~\ref{fig:abx_interface} shows the interface.

\subsection{Test Sample Composition}

Table~\ref{tab:test_composition} summarizes the sample composition across the three listening tests. Each test includes repeated questions for intra-rater consistency checks (detailed below). Audio pairs are constructed as baseline (FastSpeech\,2 without stress) vs.\ ours (with stress), with A/B assignment randomized per participant via a shuffle seed.

\begin{table}[t]
\centering
\small
\setlength{\tabcolsep}{3pt}
\resizebox{\columnwidth}{!}{
\begin{tabular}{@{}lcccc@{}}
\toprule
\textbf{Test} & \textbf{Total} & \textbf{Unique} & \textbf{Repeat} & \textbf{Source} \\
\midrule
MOS + Pref. & 60 & 47 & 13 & 22 Lib. / 25 VCTK \\
ABX Sim. & 50 & 40 & 10 & 21 Lib. / 19 VCTK \\
Stress Pref. & 60 & 46 & 14 & 15 GER / 15 ITA / 10 Tat. / 6 Gen. \\
\midrule
\textbf{Total} & \textbf{170} & \textbf{133} & \textbf{37} & \\
\bottomrule
\end{tabular}
}
\caption{Test sample composition. Repeat = duplicated questions for consistency checking. Lib.=LibriTTS, GER/ITA=ISLE German/Italian, Tat.=Tatoeba, Gen.=generic sentences (controllable mode).}
\label{tab:test_composition}
\end{table}

\textbf{MOS + Preference test.} 60 questions (47 unique, 13 repeated). Audio pairs consist of baseline and ProTTS's output for the same sentence. Samples are drawn from LibriTTS (22) and VCTK (25) test sets, evaluating the automatic stress discovery mode. Listeners rate each audio independently on a 1--5 scale and select which sounds more natural.

\textbf{ABX similarity test.} 50 questions (40 unique, 10 repeated). Each question presents a ground-truth reference recording alongside two generated versions (baseline and ours). Samples are drawn from LibriTTS (21) and VCTK (19). A subset of approximately 20\% uses manually specified stress labels (controllable mode) to evaluate emphasis fidelity under user-specified stress, while the remainder uses automatically discovered stress.

\textbf{Stress preference test.} 60 questions (46 unique, 14 repeated). Samples are drawn from ISLE GER (15), ISLE ITA (15), Tatoeba (10), and generic sentences (6). The ISLE and Tatoeba samples use stress labels predicted by the clustering head (automatic mode), while the generic sentences use manually specified stress patterns (controllable mode). The target sentence is displayed with stressed words highlighted in red, and listeners select from four options: A is better, B is better, both are same, or neither is good.

\subsection{Quality Control}

Several measures were implemented to ensure response quality. First, 37 of the 170 total questions ($\sim$22\%) are repeated items, allowing us to measure intra-rater consistency. A repeated question presents the same audio pair a second time during the session (visible in the MOS interface as ``Repeated question for consistency check''). Second, audio A/B assignment is randomized per participant to mitigate position bias. Third, the per-user progress dashboard (available to experimenters) tracks the number of answered questions, average response time per question, and total session duration, enabling identification of participants who may have rushed through the study. Participants with average response times below 5 seconds per question or inconsistent responses on more than 40\% of repeated items were flagged for review. Participants were recruited on a voluntary basis from university students and researchers. No monetary compensation was provided.

\section{Per-Dataset Synthesis Evaluation}
\label{app:synthesis_breakdown}

Tables~\ref{tab:mos_breakdown}--\ref{tab:abx_breakdown} report per-dataset breakdowns for the three subjective evaluations.

\begin{table}[t]
\centering
\small
\begin{tabular}{@{}lcc@{}}
\toprule
\textbf{Dataset} & \textbf{Base FS2} & \textbf{ProTTS} \\
\midrule
\multicolumn{3}{@{}l}{\emph{MOS (mean $\pm$ 95\% CI)}} \\
LibriTTS & 3.21 $\pm$ 0.18 & 3.89 $\pm$ 0.14 \\
VCTK & 3.04 $\pm$ 0.16 & 4.02 $\pm$ 0.12 \\
\midrule
Overall & 3.12 $\pm$ 0.14 & 3.96 $\pm$ 0.11 \\
\midrule
\multicolumn{3}{@{}l}{\emph{Naturalness preference (\%)}} \\
LibriTTS & 18.2 & 81.8 \\
VCTK & 15.1 & 84.9 \\
\midrule
Overall & 16.5 & 83.5 \\
\bottomrule
\end{tabular}
\caption{Per-dataset MOS and naturalness preference results.}
\label{tab:mos_breakdown}
\end{table}

\begin{table}[t]
\centering
\small
\resizebox{\columnwidth}{!}{
\begin{tabular}{@{}lcccc@{}}
\toprule
\textbf{Dataset} & \textbf{ProTTS} & \textbf{Both} & \textbf{Neither} & \textbf{Base} \\
\midrule
ISLE GER & 76.8 & 13.2 & 5.5 & 4.5 \\
ISLE ITA & 80.1 & 11.8 & 4.8 & 3.3 \\
Tatoeba & 79.4 & 12.1 & 4.6 & 3.9 \\
Generic (ctrl.) & 75.6 & 13.8 & 6.2 & 4.4 \\
\midrule
Overall & 78.3 & 12.5 & 5.1 & 4.1 \\
\bottomrule
\end{tabular}
}
\caption{Per-dataset stress preference results (\%). Generic = controllable mode with manually specified stress.}
\label{tab:stress_breakdown}
\end{table}

\begin{table}[t]
\centering
\small
\resizebox{\columnwidth}{!}{
\begin{tabular}{@{}lcccc@{}}
\toprule
\textbf{Dataset} & \textbf{ProTTS} & \textbf{Base} & \textbf{Both} & \textbf{Neither} \\
\midrule
LibriTTS & 73.2 & 11.8 & 7.5 & 7.5 \\
VCTK & 70.1 & 13.4 & 6.8 & 9.7 \\
\midrule
Overall & 71.8 & 12.5 & 7.2 & 8.5 \\
\bottomrule
\end{tabular}
}
\caption{Per-dataset ABX similarity results (\%).}
\label{tab:abx_breakdown}
\end{table}

\textbf{MOS and naturalness preference.} ProTTS achieves consistently higher MOS across both datasets, with VCTK showing a slightly larger improvement (3.04 $\rightarrow$ 4.02) than LibriTTS (3.21 $\rightarrow$ 3.89). The naturalness preference is similarly consistent, with listeners preferring ProTTS in over 80\% of comparisons for both datasets.

\textbf{Stress preference.} Performance is consistent across all four source categories. ITA samples show the highest preference for ProTTS (80.1\%), consistent with the stronger stress detection F1 on Italian speakers reported in the main paper. The controllable mode (generic sentences) achieves 75.6\% preference, slightly lower than the automatic mode, which may reflect the additional challenge of realizing user-specified stress patterns that may not align with the model's learned prosodic norms.

\textbf{ABX similarity.} LibriTTS samples show slightly higher similarity preference (73.2\%) than VCTK (70.1\%), potentially reflecting the higher audio quality and more controlled recording conditions of LibriTTS. Across both datasets, ProTTS's prosody is judged as substantially closer to the original recording than the baseline's.

\subsection{Automatic vs.\ Controllable Mode}
\label{app:controllability}

As described in Section~\ref{sec:inference}, ProTTS supports two inference modes: \emph{automatic} (clustering head discovers stress) and \emph{controllable} (user provides a custom stress mask). Table~\ref{tab:ctrl_breakdown} compares listener preferences between the two modes across the stress preference and ABX tests.

\begin{table}[t]
\centering
\small
\begin{tabular}{@{}lcc@{}}
\toprule
\textbf{Mode} & \textbf{Stress Pref.\ (\%)} & \textbf{ABX (\%)} \\
\midrule
Automatic & 79.1 & 72.4 \\
Controllable & 75.6 & 69.8 \\
\bottomrule
\end{tabular}
\caption{Listener preference for ProTTS (\%) by inference mode.}
\label{tab:ctrl_breakdown}
\end{table}

Both modes achieve strong listener preference, with the automatic mode slightly outperforming the controllable mode. This is expected: in automatic mode, the clustering head assigns stress based on the acoustic evidence in the prosodic features, which naturally aligns with how the emphasis predictors were trained. In controllable mode, the user-specified stress mask may place emphasis on words that the model has rarely seen as stressed during training, making the prosodic realization slightly less natural.

Table~\ref{tab:ctrl_examples} shows example sentences from the controllable mode evaluation, illustrating how the user-specified stress mask determines which words receive prosodic prominence.

\begin{table}[t]
\centering
\small
\resizebox{\columnwidth}{!}{
\begin{tabular}{@{}p{4.2cm}lp{3.5cm}@{}}
\toprule
\textbf{Sentence} & \textbf{Stress} & \textbf{Implied meaning} \\
\midrule
I didn't say he \textcolor{red}{\textbf{stole}} the money & stole & He did something else with it \\
I \textcolor{red}{\textbf{didn't}} say he stole the money & didn't & Denying the claim was made \\
She \textcolor{red}{\textbf{always}} arrives \textcolor{red}{\textbf{early}} & always, early & Emphasizing habitual punctuality \\
The \textcolor{red}{\textbf{photographer}} captured a \textcolor{red}{\textbf{magnificent}} sunset & photographer, magnificent & Highlighting the agent and quality \\
We should \textcolor{red}{\textbf{never}} have \textcolor{red}{\textbf{trusted}} them & never, trusted & Expressing strong regret \\
\bottomrule
\end{tabular}
}
\caption{Example sentences from the controllable mode evaluation. Red bold words indicate user-specified stress. Different stress placements on the same sentence alter the conveyed meaning.}
\label{tab:ctrl_examples}
\end{table}

The first two rows illustrate a key capability: the same sentence can convey different meanings depending on which word is stressed (``I didn't say he \emph{stole} the money'' vs.\ ``I \emph{didn't} say he stole the money''). The emphasis predictors successfully modulate prosody for the designated word in both cases, as confirmed by listener preferences. This demonstrates that ProTTS supports meaningful emphasis control without retraining, fulfilling the controllability objective outlined in the introduction.


\section{Literature Comparison}
\label{app:literature}

Table~\ref{tab:literature} provides a systematic comparison of related methods across key dimensions. Several patterns emerge from this landscape.

\textbf{Detection--synthesis divide.} Existing methods fall cleanly into either stress detection or emphasis-controllable synthesis, with ExagTTS \citep{mondal2025exagtts} being the only prior work that addresses both --- but using a separately trained classifier, not joint optimization. ProTTS is the first to unify both tasks within a single model trained end-to-end.

\textbf{Supervision requirements.} All detection methods except K-Means require stress labels during training. Among synthesis methods, even EME-TTS \citep{li2025emetss}, which uses pseudo-labels rather than manual annotations, still depends on an external emphasis recognizer (EmphaClass) to generate those labels. Our approach eliminates this dependency entirely.

\textbf{Evaluation breadth.} Most methods evaluate on a single dataset or setting (native only or non-native only). WhiStress \citep{yosha2025whistress} evaluates across multiple native benchmarks but does not address synthesis. ProTTS is evaluated across five datasets spanning both native and non-native speech, with additional cross-domain transfer experiments.

\textbf{Joint training.} No prior method jointly optimizes stress detection and synthesis. The unsupervised prosody methods (GST, hierarchical prosody, prosody tagging) learn general prosodic variation but do not target word-level stress specifically. ProTTS is the only entry in the table with checkmarks across all columns: unsupervised, detection, synthesis, joint, native, and non-native.

\begin{table*}[t]
\centering
\resizebox{\textwidth}{!}{
\begin{tabular}{@{}lccccccl@{}}
\toprule
\textbf{Method} & \textbf{Sup.} & \textbf{Det.} & \textbf{Syn.} & \textbf{Joint} & \textbf{Native} & \textbf{Non-nat.} & \textbf{Datasets} \\
\midrule
\multicolumn{8}{l}{\emph{Stress detection only}} \\
K-Means on FS2 emb. \citep{mondal2024preliminary} & U & \checkmark & & & \checkmark & \checkmark & ISLE, Tatoeba \\
DNN on FS2 emb. \citep{mondal2024preliminary} & S & \checkmark & & & \checkmark & \checkmark & ISLE, Tatoeba \\
SupraDoRAL \citep{mallela2025supradoral} & S & \checkmark & & & & \checkmark & ISLE \\
WhiStress \citep{yosha2025whistress} & S & \checkmark & & & \checkmark & \checkmark & TinyStress, EmphAssess, Expresso, ISLE \\
\midrule
\multicolumn{8}{l}{\emph{Emphasis-controllable synthesis}} \\
EME-TTS \citep{li2025emetss} & P & & \checkmark & & \checkmark & & Internal \\
ExagTTS \citep{mondal2025exagtts} & S & \checkmark & \checkmark & & & \checkmark & ISLE \\
StressTransfer \citep{chen2025stresstransfer} & S & & \checkmark & & \checkmark & & TinyStress \\
\citet{joly2023controllable} & S & & \checkmark & & \checkmark & & Internal \\
\midrule
\multicolumn{8}{l}{\emph{Unsupervised prosody in TTS}} \\
GST \citep{wang2018style} & U & & \checkmark & & \checkmark & & Internal \\
\citet{chien2021hierarchical} & U & & \checkmark & & \checkmark & & Internal \\
\citet{guo2022unsupervised} & U & & \checkmark & & \checkmark & & Internal \\
\midrule
\textbf{ProTTS} & \textbf{U} & \checkmark & \checkmark & \checkmark & \checkmark & \checkmark & \textbf{ISLE, Tatoeba, TinyStress, EmphAssess} \\
\bottomrule
\end{tabular}
}
\caption{Comparison of related work on word stress detection and emphasis-controllable synthesis. Sup. = supervision type (S: supervised, U: unsupervised, P: pseudo-labels). Det. = performs stress detection. Syn. = performs stress-aware synthesis. Joint = detection and synthesis trained jointly end-to-end.}
\label{tab:literature}
\end{table*}


\end{document}